\documentclass[aps,prb,twocolumn,showpacs,letterpaper,tighten,float,superscriptaddress,nofootinbib,reprint]{revtex4-2}
\usepackage{amssymb}
\usepackage{amsbsy}
\usepackage{amsmath}
\usepackage{mathrsfs}
\usepackage{epsfig}
\usepackage{graphicx}
\usepackage{array}
\usepackage{textcomp}
\usepackage{color}
\usepackage{braket}
\usepackage{hhline}
\usepackage{comment}
\usepackage{bm,hyperref}
\usepackage[justification=raggedright,font=small]{caption}
\usepackage[titletoc,toc,title]{appendix}
\usepackage[normalem]{ulem}
\usepackage[labelformat=simple]{subcaption}

\usepackage{array}
\newcolumntype{P}[1]{>{\centering\arraybackslash}p{#1}}
\newcolumntype{M}[1]{>{\centering\arraybackslash}m{#1}}

\definecolor{myblue}{rgb}{.93, .93, 1}

\definecolor{darkgreen}{rgb}{0,0.7,0}

\newcommand{\beq}{\begin{equation}}
\newcommand{\eeq}{\end{equation}}

\newcommand{\bpm}{\begin{pmatrix}}
\newcommand{\epm}{\end{pmatrix}}
\newcommand{\bmm}{\begin{matrix}}
\newcommand{\emm}{\end{matrix}}

\graphicspath{{./Figures/}}

\begin{document}

\author{Abhinav Prem}
\affiliation{
Department of Physics,
Princeton University, NJ 08544, USA
}
\affiliation{School of Natural Sciences, Institute for Advanced Study, Princeton, NJ 08540, USA}
\author{Vir B. Bulchandani}
\affiliation{
Department of Physics,
Princeton University, NJ 08544, USA
}
\affiliation{
Princeton Center for Theoretical Science,
Princeton University, NJ 08544, USA
}
\author{S. L. Sondhi}
\affiliation{Rudolf Peierls Centre for Theoretical Physics, University of Oxford, Oxford OX1 3PU, United Kingdom}
\date{\today}

\title{Dynamics and transport in the boundary-driven dissipative Klein-Gordon chain}
\date{\today}

\begin{abstract}
Motivated by experiments on chains of superconducting qubits, we consider the dynamics of a classical Klein-Gordon chain coupled to coherent driving and subject to dissipation solely at its boundaries. As the strength of the boundary driving is increased, this minimal classical model recovers the main features of the ``dissipative phase transition'' seen experimentally. Between the transmitting and non-transmitting regimes on either side of this transition (which support ballistic and diffusive energy transport respectively), we observe additional dynamical regimes of interest. These include a regime of superdiffusive energy transport at weaker driving strengths, together with a ``resonant nonlinear wave'' regime at stronger driving strengths, which is characterized by emergent translation symmetry, ballistic energy transport, and coherent oscillations of a nonlinear normal mode. We propose a non-local Lyapunov exponent as an experimentally measurable diagnostic of many-body chaos in this system, and more generally in open systems that are only coupled to an environment at their boundaries.
\end{abstract}

\maketitle


\section{Introduction}
\label{sec:intro}

The relaxation dynamics of a many-body system at long times is generically dominated by a small number of hydrodynamic slow modes, corresponding to its local conserved charges. A classic method for probing such hydrodynamic behavior consists of coupling the boundaries of the system to thermal reservoirs at different temperatures~\cite{dhar2019}, thereby forcing energy flow through the system, and waiting for a locally equilibrated steady state to form in the bulk. Coupling to the boundaries of a system in this way can be viewed as a ``weak'' perturbation, in the sense that it allows one to probe bulk transport in the non-equilibrium steady state (NESS) without altering the properties of the system. This technique has been used to elucidate transport in both ergodic and non-ergodic classical~\cite{dhar2019} and quantum~\cite{Bertini_2021} many-body systems. 

Recent experiments on coupled arrays of circuit QED resonators~\cite{fitzpatrick2017,fedorov2021} explore a different kind of boundary driving, whereby a given system is driven \textit{coherently} at one end by an external laser and the outgoing radiation is measured at the other end. This interplay between coherent driving and incoherent dissipation, induced by the intrinsic loss rates of photonic cavities, opens new avenues for exploring genuinely non-equilibrium steady states and the possible phase transitions between them. The effects of such coherent boundary driving are far less well-studied theoretically than the effects of coupling to incoherent thermal reservoirs; we refer to Refs.~\cite{Johansson_2009,biella2015,debnath2017,eckmann2020,etOlla} for some recent investigations. For a generic, thermalizing system that is driven in this manner, one might expect that at long times the system again approaches a local equilibrium state, which is characterized by hydrodynamic behavior and which interpolates smoothly between an effective temperature set by the coherent drive and zero effective temperature at the non-driven end. However, this expectation is only reasonable at low driving frequencies and amplitudes. Continuous driving of the system at large frequencies will in general sustain non-hydrodynamic degrees of freedom, while driving at large enough amplitudes will eventually hamper local equilibration. This raises the question of which aspects of a bulk system's physics dictate its response to strong, coherent boundary driving. 

\begin{figure*}
    \centering
    \includegraphics[width=\textwidth]{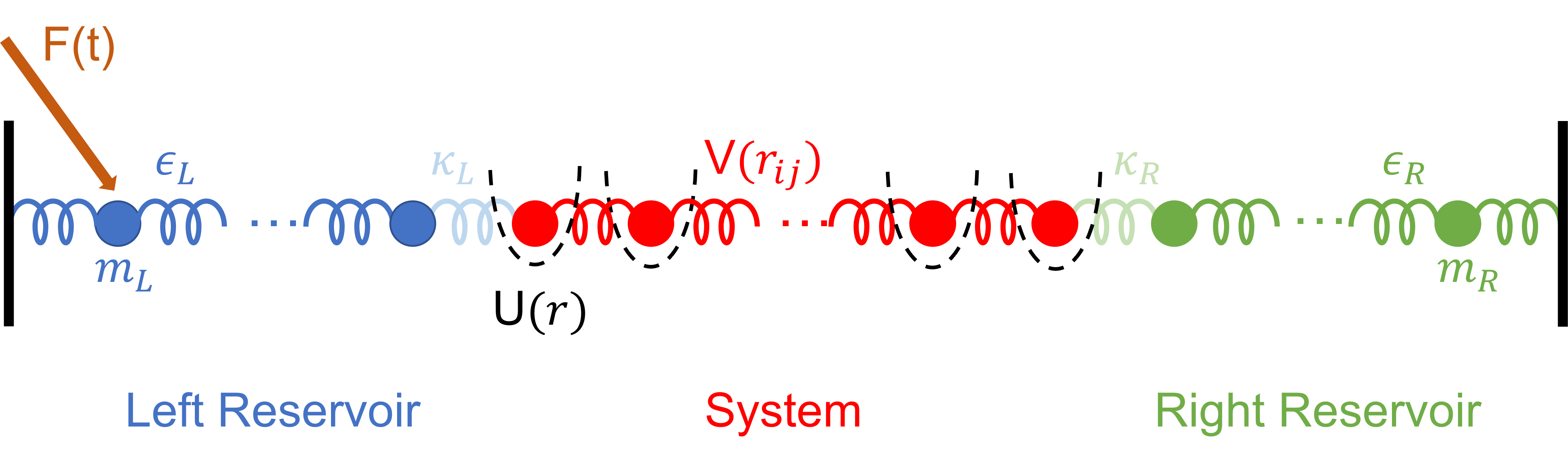}
    \caption{Schematic of the set-up considered in this paper. The system consists of $N$ anharmonic oscillators with on-site potential $U(r)$ and inter-particle coupling $V(r_{ij})$. The system is coupled at either end to two reservoirs, which are taken to be semi-infinite harmonic chains. The left hand bath is also subjected to an external driving force $F(t)$.}
    \label{fig:setup}
\end{figure*}

Here, we study this question systematically in a classical Klein-Gordon chain which is coherently driven at one end and is subject to dissipation at both of its boundaries. Our choice of model is particularly motivated by the superconducting circuit experiments of Ref.~\cite{fitzpatrick2017,fedorov2021}, which explored the nonequilibrium transport of interacting photons in a boundary-driven setting. We focus on dissipation that is spatially localized at the ends of the chain because the dominant source of dissipation in these experimental systems is typically set by the photon loss rates of the first and last cavities~\cite{fedorov2021}. These experiments observed a transition between ``transmitting'' and ``non-transmitting'' regimes as the driving strength was increased,
which was interpreted as a ``dissipative phase transition'' and was reproduced within a semiclassical analysis~\cite{fitzpatrick2017}. Qualitatively similar features were also observed numerically in the semiclassical regime of the boundary-driven Bose-Hubbard model~\cite{debnath2017}, which suggests that such dynamical behavior should arise even in boundary-driven dissipative \textit{classical} nonlinear systems. 

Our results on the boundary-driven dissipative Klein-Gordon chain show that the coherent injection of energy provided by a monochromatic external drive can lead to both conventional hydrodynamic behavior and to far-from-equilibrium phases that are beyond the purview of conventional theories of hydrodynamics and transport. For example, while we find that the experimentally observed dynamical regimes of high and low transmission are recovered within our model at weak and strong driving amplitudes respectively, with the weakly-driven regime exhibiting conventional near-harmonic behavior, the strongly-driven regime exhibits an unusual coexistence of diffusive energy transport with incomplete local thermalization. We also uncover additional dynamical regimes at intermediate driving strengths, which were not seen experimentally; these include a locally equilibrated regime that hosts conventional hydrodynamic behavior, albeit with an anomalous dynamical exponent, and an unconventional, non-hydrodynamic \textit{resonant nonlinear wave} regime that shows both ballistic energy transport and coherent nonlinear oscillations at the driving frequency. All regimes but the latter persist up to the largest system sizes we have investigated numerically, which suggests that they correspond to genuine non-equilibrium phases of matter\footnote{To show convincingly that these distinct regimes survive in the thermodynamic limit would require theoretical arguments beyond the primarily numerical evidence in this paper. For this reason, we refer to dynamical ``regimes" rather than ``phases" throughout this work.}. In contrast, our numerics indicate that the resonant nonlinear wave regime is absent in the thermodynamic limit, but should nevertheless appear as a distinct dynamical regime at system sizes accessible to current experiments. These dynamical regimes and their qualitative features are summarized in Table~\ref{table}.

Finally, besides diagnosing thermalization (or the lack thereof) in these distinct NESSs, we propose a \textit{non-local Lyapunov exponent} as a practically useful diagnostic for chaos in the bulk of the system. This yields a minimally disruptive method for probing the bulk dynamics and involves weakly perturbing the phase of the external drive at one end of the system, before studying the effect of this small change on the radiation emitted at the opposite end of the system. 
\begin{table}[b]
{\renewcommand{\arraystretch}{1.5}
\begin{tabular}{ |P{2.1cm}|P{2.1cm}|P{2.1cm}|P{2.1cm}|  }
 \hline
Quasilinear Regime & Thermal Regime & Resonant Nonlinear Wave Regime & Partially Thermal Regime\\
 \hline\hline
  Non-thermal NESS in the bulk & Thermal NESS in the bulk & Non-thermal NESS in the bulk & Thermal NESS in a subregion \\
 \hline
 $j \sim F_d^2$ & $j \sim F_d^{\alpha (>2)}$ & $j \sim F_d^0$ &  $j \sim F_d^{-2}$ \\
 \hline
Ballistic $j\sim N^0$ & Superdiffusive $j \sim N^{-0.40}$ & Ballistic $j \sim N^0$ & Diffusive $j \sim N^{-1}$ \\
 \hline
  Regular dynamics ($\lambda = 0$) & Chaotic dynamics ($\lambda > 0$) & Regular dynamics ($\lambda = 0$) & Chaotic dynamics ($\lambda > 0$)  \\
 \hline
\end{tabular}
\caption{Dynamical regimes for the boundary-driven dissipative Klein-Gordon chain at system sizes $N\sim 500$. The distinct regimes (going left across the table) are obtained by increasing the driving amplitude $F_d$ at fixed driving frequency $\omega_d$. Here, $j$ refers to the steady-state current and $\lambda$ to the non-local Lyapunov exponent (see Sec.~\ref{sec:probes}).}
\label{table}
}
\end{table}

This paper is organized as follows: in Sec.~\ref{sec:models}, we define the boundary-driven dissipative Klein-Gordon chain starting from a microscopic model that includes the coupling to external reservoirs and an external coherent drive. As we show in Appendix~\ref{sec:eliminate}, the reservoirs can be systematically eliminated with the effective dynamics of the system governed by generalized Langevin equations that include a coherent drive acting on one end and incoherent dissipation localized at both ends of the chain. In Sec.~\ref{sec:probes}, we discuss the various probes, including the energy current and the non-local Lyapunov exponent, that we use to diagnose thermalization and chaos in the NESS we obtain. As a warm-up, we consider the analytically tractable case of the boundary-driven dissipative harmonic chain in Sec.~\ref{sec:linear}. We present the main results of this paper on the fully anharmonic chain in Sec.~\ref{sec:kg}. The distinct dynamical regimes that we find for this model are characterized in Secs.~\ref{sec:weak}-\ref{sec:mixed}, while results on the non-local Lyapunov exponent are presented in Sec.~\ref{sec:lyapunov}. We conclude in Sec.~\ref{sec:cncls} with a summary of results and a discussion of open questions.


\section{Model and General Setup}
\label{sec:models}

In this paper, we consider the setup shown in Fig.~\ref{fig:setup}, which is motivated by current circuit QED experiments wherein a one-dimensional array of coupled circuit elements is connected via its boundaries to external reservoirs, which can themselves be coherently driven. This setup generalizes the Rubin model of dissipation~\cite{rubin1960,rubin1963,weiss2012quantum}, which consists of a single impurity embedded in an infinite one-dimensional harmonic chain and provides a simple physical model of a heat bath whose spectral properties can be analytically calculated in terms of the microscopic parameters of the harmonic chain. Instead of a single impurity, we consider a chain of finitely many anharmonic oscillators linearly coupled at its ends to leads, a setup which has previously been considered in the context of heat conduction~\cite{eckmann1999,das2012,dhar2012,wang2014}. Here, however, we are interested instead in the interplay between driving and dissipation on the system's long time steady-state. We thus add an external time-dependent driving force to the left bath which mediates an effective driving on the system. Finally, while we model the reservoirs as Rubin baths, we will set their temperatures to zero such that they exert a vanishing random force on the system and only generate damping at the boundaries.

The system consists of $N$ nonlinear oscillators with positions $\{q_j\}_{j=1}^N$ and momenta $\{p_j\}_{j=1}^N$. The degrees of freedom of the baths are denoted by $\{q_{j,\alpha},p_{j,\alpha}\}_{j=1}^{N_\alpha}$, where $\alpha = L,R$ and $N_\alpha$ specifies the number of reservoir oscillators. The classical dynamics of the coupled system and reservoirs is governed by the following time-dependent Hamiltonian:
\beq
\label{eq:totalH}
H = H_S + H_{B,L} + H_{B,R} + H_{c,L} + H_{c,R} + H_d(t) \, ,
\eeq
where the system Hamiltonian is given by
\begin{align}
\label{eq:sysHam}
H_S = & \sum_{j=1}^N \left[ \frac{p_j^2}{2m} + U(q_j) \right] + \sum_{j=1}^{N-1} V\left(q_{j+1} - q_j \right) \nonumber \\
+ & \frac{\kappa_L}{2} q_1^2 + \frac{\kappa_R}{2} q_N^2 \, .
\end{align}
Here, $m$ denotes the mass of the system's oscillators while $U(r)$ and $V(r_{ij})$ specify the on-site and interaction potentials respectively. Note that the final two terms come from the coupling between the first (last) oscillator of the system and the last (first) oscillator of the left (right) baths, with $\kappa_\alpha$ denoting the respective coupling strengths. Conventionally, one also includes a counter-term~\cite{caldeira81,caldeira83} to ensure that the system-bath coupling does not renormalize the bare on-site potential seen by the system's degrees of freedom; however, as this renormalization does not arise in the $N_{\alpha} \to \infty$ limit, we omit such counter-terms here. 

\begin{figure}[t]
    \centering
    \includegraphics[width=0.5\textwidth]{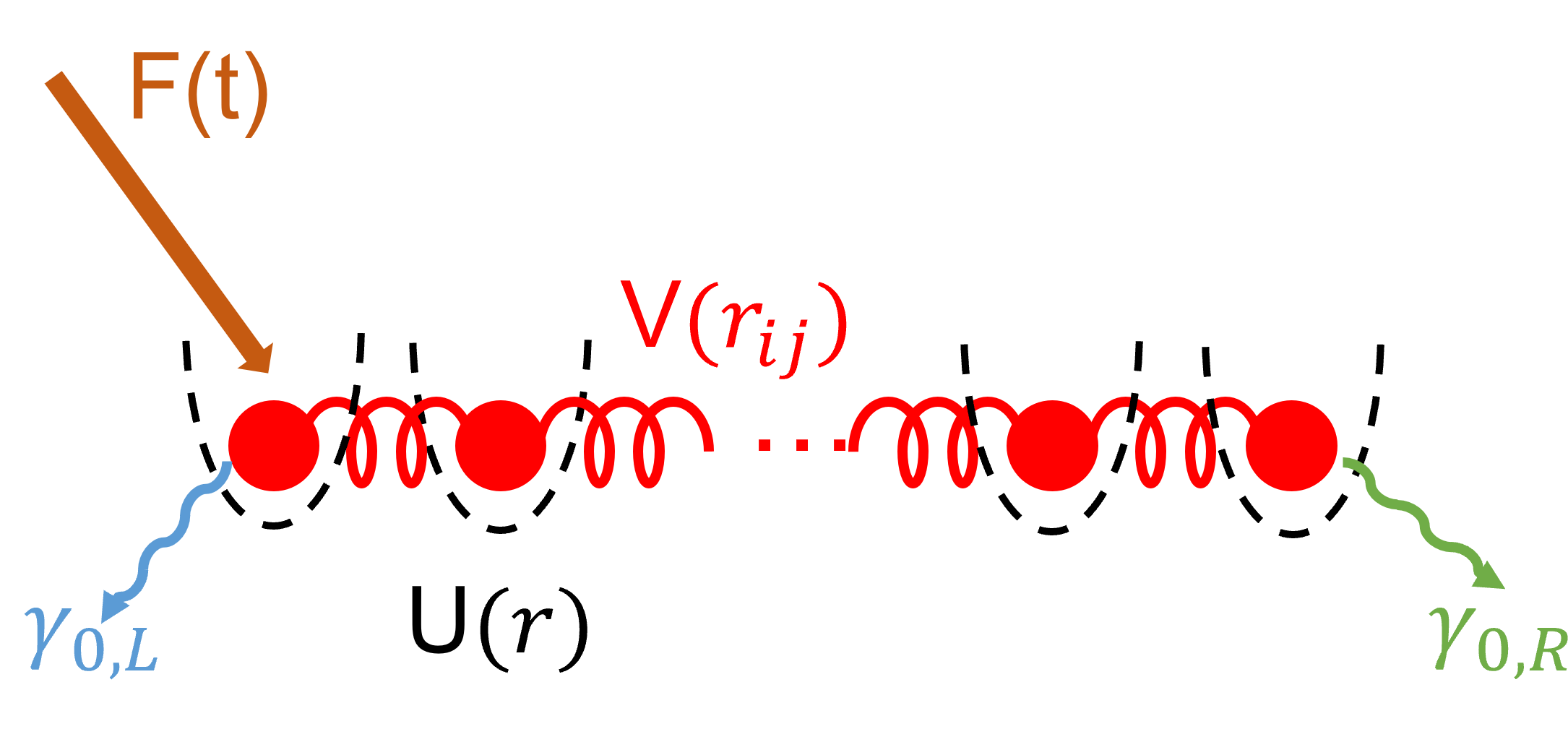}
    \caption{Effective description of the system after eliminating the reservoirs. The system experiences boundary driving and dissipation. }
    \label{fig:setup2}
\end{figure}

The bath Hamiltonians are given by
\begin{align}
H_{B,L} &= \sum_{j=1}^{N_L} \left[ \frac{p_{j,L}^2}{2 m_L} + \frac{\epsilon_L}{2} \left(q_{j,L} - q_{j-1,L} \right)^2 \right] + \frac{\kappa_L}{2} q_{N_L,L}^2  \, ,\\
H_{B,R} &= \sum_{j=1}^{N_R} \left[ \frac{p_{j,R}^2}{2 m_R} + \frac{\epsilon_R}{2} \left(q_{j+1,R} - q_{j,R} \right)^2 \right] + \frac{\kappa_R}{2} q_{1,R}^2 \, ,
\end{align}
where we fix the left (right) boundary of the left (right) bath to be fixed $q_{0,L} = q_{N+1,R} = 0$. The mass and inter-particle spring constants of the reservoir oscillators are given by $m_\alpha$ and $\epsilon_\alpha$ respectively. The terms describing the linear coupling between the system and reservoirs are given by
\beq
H_{c,L} = - \kappa_L \, q_1 \, q_{N_L,L} , \quad H_{c,R} = -\kappa_R \, q_N \, q_{1,R}  \, , 
\eeq
while the external driving is included through the term 
\beq
H_d(t) = -q_{L,1} F(t) \, . 
\eeq

In Appendix~\ref{sec:eliminate}, we systematically eliminate the reservoir degrees of freedom to derive the dynamical equations for the system. Specifically, we assume a continuous distribution of bath modes ($N_\alpha \to \infty$), set the temperatures of the reservoirs to zero, and take the continuum string limit, which results in an Ohmic spectral density for the reservoirs. This approximation results in Markovian dissipation that acts only on the first and last sites of the system and an external coherent force that drives the first site (see Fig.~\ref{fig:setup2}); the resulting equations of motion are:
\begin{align}
    \dot{p}_1 &= - U'(q_1) + V'(q_2 -q_1) -\gamma_{0,L} \dot{q}_1(t) + 2 F(t) \,   \label{eq:LE1} \\ 
    \dot{p}_j &=  -U'(q_j) + V'(q_{j+1} - q_j) - V'(q_j - q_{j-1}) \,  \\
    \dot{p}_N& = - U'(q_N) - V'(q_N -q_{N-1}) - \gamma_{0,R} \dot{q}_N(t) \, , \label{eq:LEn} 
\end{align}
with the system initialised at rest: $q_j(0) = p_j(0) = 0$.

In the remainder of this paper, we will study the non-equilibrium steady states of the (finite) inhomogeneous boundary-driven dissipative system described by Eqs.~\eqref{eq:LE1}-\eqref{eq:LEn}. We will consider the following on-site and inter-particle couplings:
\beq
U(r) = \frac{m \omega^2}{2} r^2 + \nu \frac{\lambda}{4} r^4 \, , \quad V(r_{ij}) = \frac{\epsilon}{2} r_{ij}^2 \, ,
\eeq
where $\nu \in \{0,1\}$. For $\nu = 0$, these potentials describe a harmonic chain, while for $\nu = 1$ they describe an anharmonic Klein-Gordon chain with both a quadratic and a quartic on-site potential. Further, we assume symmetric boundary dissipation $\gamma_{0,\alpha} = \eta$ and take the external driving to be periodic: $F(t) = F_d/2 \cos(\omega_d t)$. Setting $m=1$, we can further reduce the number of independent scaling parameters by suitably re-scaling
\beq
q_j \to \sqrt{\frac{\lambda}{\omega^2}} q_j, \quad t \to \omega t \, ,
\eeq
such that Eqs.~\eqref{eq:LE1}-\eqref{eq:LEn} become
\begin{align}
    \ddot{q}_1 &= - q_1 - \nu q_1^3 - \tilde{\epsilon}(q_1 - q_2) -\tilde{\eta} \dot{q}_1 + \tilde{F}_d \cos(\tilde{\omega}_d t) \, \label{eq:maineq1}  \\ 
    \ddot{q}_j &=  -q_j - \nu q_j^3 - \tilde{\epsilon} (2q_j - q_{j+1} - q_{j-1}) \,  \\
    \ddot{q}_N& = - q_N - \nu q_N^3 - \tilde{\epsilon}(q_N-q_{N-1}) - \tilde{\eta} \dot{q}_N \, ,  \label{eq:maineqN}
\end{align}
where the re-scaled parameters are
\beq
\label{eq:relabel}
\tilde{\epsilon} = \frac{\epsilon}{\omega^2}, \, \tilde{\eta} = \frac{\eta}{\omega}, \, \tilde{F}_d = \frac{\sqrt{\lambda}}{\omega^3}F_d, \, \tilde{\omega}_d = \frac{\omega_d}{\omega} \, .
\eeq
We henceforth drop the tildes to reduce clutter and note that the strength of the nonlinearity now appears through the re-scaled driving amplitude. While this defines a rather large and rich parameter space, we will primarily study this system of equations as a function of the driving amplitude $F_d$ and driving frequency $\omega_d$, as well as the system size $N$. Since the external drive parameters $\omega_d$ and $F_d$ are likely the easiest to tune in any experiment, we are most interested in the distinct dynamical regimes that can appear at a fixed system size as the drive parameters are varied.

The Hamiltonian corresponding to the above equations of motion is given by
\beq
\label{eq:HSys}
H_S(\nu) = \sum_{j=1}^N \left[\frac{p_j^2}{2} +\frac{q_j^2}{2}  + \nu \frac{q_j^4}{4}\right] + \frac{\epsilon}{2}\sum_{j=1}^{N-1} (q_{j+1} - q_j)^2 \,
\eeq
where $\nu = 0$ (resp. 1) corresponds to the harmonic (resp. anharmonic) chain and should not be confused with the strength of the nonlinearity, which we set to unity in the anharmonic case. The external forcing can be accounted for through the time-dependent term: $H_d(t) = q_1 F_d \cos(\omega_d t)$. Note that the driving term explicitly breaks translation invariance, leading to all modes of the underlying harmonic lattice being driven. 

\section{Diagnostics for Chaos and Thermalization}
\label{sec:probes}

In this section, we discuss the quantities we will use to diagnose ergodicity and thermalization (or lack thereof) in the boundary-driven dissipative Klein-Gordon chain. \\

\paragraph*{Local temperature and energy current: } To explore the dynamics of the boundary-driven dissipative anharmonic chain, we will focus mostly on the transport properties in the NESS induced by the interplay between the driving and dissipation; specifically, we study the local temperature profiles and the average energy current in the NESS. Defining a notion of local temperature first requires defining a notion of a local equilibrium state. Representing the Hamiltonian $H_S$ in Eq.~\eqref{eq:HSys} as a sum of local terms $H_S = \sum_{j=1}^N h_j$, a natural class of local (canonical) equilibrium states for this system comprises phase space measures of the form
\begin{equation}
\label{eq:loceqstate}
\rho(p,q) = \frac{1}{Z} \exp{\left(-\sum_{j=1}^N \beta_j h_j(p,q)\right)}
\end{equation}
where $Z$ denotes the appropriate normalization constant and $\beta_j$ corresponds to the local inverse temperature. Assuming Eq.~\eqref{eq:loceqstate} is a valid description of the NESS (below we will test this assumption numerically), an integration by parts implies a local version of the equipartition theorem~\cite{toda2012statistical}
\beq
\Big \langle p_j \frac{\partial h_j}{\partial p_j} \Big \rangle = T_j,
\eeq
which allows us to define a \textit{local} temperature in terms of the second moment of the canonical momentum distribution
\beq
T_j = \langle p_j^2 \rangle \, ,
\eeq
where $\langle \cdot \rangle$ denotes the time-average in the NESS.

Next, to define the local energy current we look for a discrete version of the continuity equation
\beq
\partial_t h(x,t) + \nabla j(x,t) = 0 \, 
\eeq
in terms of the local energy density $h_\ell$. Taking the time-derivative of the local energy density $h_\ell$, we find that~\cite{leprireview,dhar2008}
\beq
\frac{\partial h_\ell}{\partial t} = - \left(j_{\ell+1,\ell} - j_{\ell,\ell-1}\right), \quad \ell\in[2,N-1]
\eeq
where
\beq
\label{eq:bcurrent}
j_{\ell,\ell-1} = -\frac{\epsilon}{2}(p_\ell + p_{\ell-1})(q_\ell - q_{\ell-1}), \quad \ell \in [2,N]
\eeq
is the energy current from site $\ell-1$ to site $\ell$. Meanwhile, for the boundary sites $\ell=1,N$, we find that
\begin{align}
    \frac{\partial h_1}{\partial t} &= -(j_{2,1} - j_{1,L}), \nonumber \\
    \frac{\partial h_N}{\partial t} &= -(j_{R,N} - j_{N,N-1}) \, ,
\end{align}
where 
\beq
j_{1,L} = p_1 F_d \cos(\omega_d t)- \eta p_1^2 \, 
\eeq
is the energy current entering the system from the left. The first term is the rate of work done by the external drive on the system and the second term gives the rate at which energy is dissipated into the left bath (the negative sign indicates the energy flux leaving site 1 into the left bath due to dissipation). Similarly, 
\beq
j_{R,N} = \eta p_N^2 \, ,
\eeq
which gives the rate at which the system dissipates energy into the right bath. The system hence attains a NESS when the average energy current $j$ is constant in time and the following equality holds
\beq
\label{eq:NESScond}
j = \langle j_{1,L} \rangle = \langle j_{\ell,\ell-1} \rangle = \langle j_{R,N} \rangle = j_B, \quad \forall \ell \in [2,N]  \, , 
\eeq
with the bulk energy current defined as 
\beq
j_{B} = \frac{1}{N-1} \sum_{l=2}^{N} \langle j_{l,l-1} \rangle \, .
\eeq
To understand the nature of transport in the anharmonic chain, we will study the scaling of the average energy current with $N$ as well as its dependence on the external drive parameters $F_d$ and $\omega_d$. 

\paragraph*{Local momentum distribution: } Another probe that reveals the nature of the non-equilibrium steady state is the canonical momentum distribution along the chain. In particular, we will check whether the local canonical momentum distributions in the NESS are Maxwell-Boltzmann with a temperature set by the local temperature $T_\ell$ at a given site $\ell$. A straightforward test revealing the (non)-Gaussianity of the canonical momentum distribution of the time-averaged ensemble (in the NESS) is given by the kurtosis (or moment-ratio test) 
\beq
\label{eq:Kurtdef}
\text{Kurt}_\ell = \frac{\braket{p_\ell^4}}{\braket{p_\ell^2}^2} \, ,
\eeq
where Kurt$_\ell = 3$ holds for a Gaussian distribution i.e., it indicates local equilibrium with respect to the thermal Maxwell-Boltzmann distribution
\beq
\label{eq:MBdist}
f(x) = \frac{1}{\sqrt{2 \pi T}} \exp\left(- \frac{x^2}{2 T} \right)\, .
\eeq
Hence, we will use the kurtosis to measure deviations from local thermal equilibrium.

\paragraph*{Normal mode distribution: } We will also find it useful to characterize the system through the Fourier distributions in the NESS, both in quasimomentum ($k$) and frequency ($\omega$) space. We use the following transformation to normal mode coordinates and canonical momenta:
\beq
\binom{q_j(t)}{p_j(t)} = \sqrt{\frac{2}{N}}\sum_{r=0}^{N-1} \binom{Q_r(t)}{P_r(t)} \cos\left[\frac{\pi(2j-1)r}{2N}\right] \, ,
\eeq
with quasimomentum $k = r \pi/N \in [0,\pi)$ ($r = 0,\dots, N-1$). Here, the $r=0$ mode corresponds to a uniform translation of the chain while $r=1,\dots,N-1$ correspond to the remaining $N-1$ vibrational modes. The above transformation differs from the standard choice because we are considering open boundary conditions. This transformation diagonalizes the undriven undamped harmonic chain, corresponding to the Hamiltonian Eq.~\eqref{eq:HSys} with $\nu = 0$: $H_S(\nu = 0) = \sum_{k}E_k$, with normal mode energies
\beq
E_k = \frac{P_k^2 + \omega_k^2 Q_k^2}{2}, \quad \omega_k = \sqrt{1 + 4 \epsilon \sin\left( \frac{k}{2}\right)^2} \, .
\eeq
We will use the time-averaged spectrum of mode energies $\braket{E_k}$ to monitor the steady-state features of the driven dissipative chain. For instance, for the harmonic chain we expect that the mode spectrum will exhibit a sharp peak (which approaches a $\delta$-function in the limit $N\to\infty$) corresponding to the mode being resonantly driven by the external force. In other words, even though the external force drives all modes (as becomes clear by transforming the driving term $H_d(t)$ to normal mode coordinates), in the thermodynamic limit only the mode resonant with the driving frequency should survive in the long-time limit, while all others decay on a shorter timescale.

\paragraph*{Power spectral density: } The frequency response of the system is probed through the quantity 
\beq
\label{eq:SWdef}
S(\omega) = \log|\mathcal{F}\{p_N^2\}(\omega)| \, ,
\eeq
where $\mathcal{F}\{\cdot\}(\omega)$ denotes the Fourier transform in time of the canonical momentum at the last site of the system. This probe is motivated by the input-output theory of open systems~\cite{breuer2007theory,rmpcqed}, which relates the correlation functions of the last site of the chain to those of the output fields i.e., of the reservoir modes, which are the experimentally measurable quantities in circuit QED architectures. In the absence of any nonlinearity in the system, we expect that the frequency response will be synchronized with the driving frequency; for the anharmonic Klein-Gordon chain, we find that the frequency response reflects ergodicity (or its failure) as well as the nature of transport in the NESS. Specifically, we find that a non-ergodic NESS with ballistic transport has a frequency response synchronized with the driving frequency, while steady states in local thermal equilibrium, which have diffusive transport, display a broadband frequency response. These results suggest that the frequency response of the output emission spectrum is a good probe for nonlinear dynamics in open systems. We note that although a broadband spectrum arises in the Klein-Gordon chain, a broadband spectrum does not imply a thermal NESS in general; for example, we expect that a nonlinear integrable system (such as the Toda lattice) will also exhibit a broadband response, despite having a nonthermal NESS.

\paragraph*{Non-local Lyapunov exponent: } As a measure of chaos~\footnote{Whether chaos is necessary for thermalization in isolated many-body classical systems is an old and yet unsettled debate~\cite{baldovin2021,salas2022}.} in the non-equilibrium steady state, we introduce the concept of a \textit{non-local Lyapunov exponent}, a diagnostic that is specifically motivated by circuit QED experiments, for reasons we explain below. Conceptually, this quantity is related to the classical version of the out-of-time-ordered correlator (OTOC)~\cite{das2018,khemani2018}, which measures how an infinitesimally localized perturbation spreads in space and grows in time. While the OTOC has been studied in various extended classical systems to gain insight into aspects of ergodicity, nonlinearity, and disorder~\cite{bilitewski2018,murugan2021,kumar2020,ruidas2020,kiran2020}, it remains largely unexplored in the context of driven-dissipative systems (see however Ref.~\onlinecite{chatterjee2020}). The OTOC is usually defined as follows: consider two identical copies ($A$ and $B$) of the system in the same initial state. At time $t=0$, locally perturb the position of the oscillator at site $\ell$ by an infinitesimal amount $\delta$ in copy $B$ and allow the systems to evolve independently under Eqs.~\eqref{eq:maineq1}-\eqref{eq:maineqN}. The OTOC
\beq
\label{eq:OTOC}
D(j,t) = \frac{|q_j^{(A)}(t) - q_j^{(B)}(t)|}{|\delta|}
\eeq
then measures the spatiotemporal spread of the initial perturbation and allows one to define a finite-time Lyapunov exponent~\cite{ftle}. 

While the above procedure is readily carried out numerically, it is practically challenging to implement in current circuit QED experiments. Spatio-temporally resolved readout of local observables (such as the local occupation number) has been achieved for a circuit QED chain with $N = 9$ sites~\cite{roushan2017}, but it is currently unfeasible for one-dimensional circuit QED arrays with several ($N \gtrsim 20$) sites to obtain local control and readout of individual lattice sites in the bulk with the requisite precision. Such experiments on large arrays typically do not directly perform local measurements in the bulk of the system but instead probe its properties indirectly by driving one end of the array with a coherent input and measuring the signal radiated at the opposite end~\cite{rmpcqed,fitzpatrick2017}. Given that these experiments have exquisite experimental control over the external driving parameters and can more straightforwardly investigate the response of the system to a change in the drive by measuring the output radiation, we propose a modified protocol that probes chaos in the bulk by measuring how the output at one boundary responds to an infinitesimal change in the input initially localized at the opposite boundary. 

Consider two copies ($A$ and $B$) of the system with identical initial conditions and allow them to evolve for several driving periods $T = 2 n \pi/\omega_d$ until they reach the steady state. At time $T$, we shift the phase of the external drive in copy $B$ by an infinitesimal amount $\phi \ll 1$ relative to copy $A$ and then allow the systems to evolve independently under Eqs.~\eqref{eq:maineq1}-\eqref{eq:maineqN}. From these equations of motion, we find that the deviation in the phase of the external drive results in a deviation $F_d/2 \, (\phi \, \delta t)^2$ between the positions of the first oscillator in the two copies at time $T + \delta t$ ($\delta t \ll 1$). In order to systematically track the spatio-temporal spread of the perturbation across the chain, we then redefine time such that $t = 0$ corresponds to the time at which the positions of the first oscillator in copies $A$ and $B$ differ by an infinitesimal amount $\delta$ ($\delta \ll \phi, \delta t \ll 1$). In other words, the redefined time $t = 0$ corresponds to 
\beq
\delta q_j(t = 0) = \delta q_j^{(B)}(t = 0) - \delta q_j^{(A)}(t = 0) = \delta \, \delta_{j,1} \, ,
\eeq
which is the time at which the infinitesimal perturbation $\phi$ in the external drive results in a perturbation $\delta$ in the position of the first oscillator. Since the experimentally relevant quantity is the output radiation, which is related to the dynamics of the last oscillator through the input-output formalism~\cite{breuer2007theory,rmpcqed}, we are interested in how this perturbation, initially localized at the first site, grows (or decays) at the last site of the chain. To capture this, we study the Lyapunov exponent, defined as
\beq
\label{eq:ftle}
\lambda_j = \lim_{t \to \infty} \lambda_j(t) = \lim_{t \to \infty} \frac{1}{t} \Big\langle \ln \frac{|\delta q_j(t)|}{|\delta|} \Big\rangle_s \, ,
\eeq
where we will focus primarily on $j = N$. Here, $\delta q_j(t) = q_j^{(A)}(t) - q_j^{(B)}(t)$ is the deviation between the two trajectories and the average $\braket{\cdot}_s$ denotes an average over initial conditions drawn from the steady-state distribution.

For a chaotic system, one expects that $\lambda_j(t)$ approaches a positive constant $\lambda_j > 0$ at late times, while for non-chaotic dynamics $\lambda_j(t)$ is always negative and saturates to some $\lambda_j \leq 0$ at late times. We note that although unbounded chaotic systems can sustain chaos for arbitrarily long times, for bounded systems (including those in local thermal equilibrium) $\lambda_j(t)$ always decays as $\sim 1/t$ at late times for a fixed perturbation strength $\delta$\footnote{Strictly speaking, one should define the Lyapunov exponent as 
\begin{equation*}
\lambda_j = \lim_{t \to \infty} \lim_{\delta \to 0} \lambda_j(t) = \lim_{t \to \infty} \lim_{\delta \to 0} \frac{1}{t}
\Big\langle \ln \frac{|\delta q_j(t)|}{|\delta|} \Big\rangle_s \, .
\end{equation*}
The limit $\delta \to 0$ is impractical both experimentally and numerically, but the exponent can nonetheless be found at fixed $\delta$ from the intermediate time behavior before the deviation saturates.} such that the Lyapunov exponent must instead be extracted from the intermediate time behavior. For a bounded chaotic system (which satisfies $|\delta q_j (t)| \leq q_j^* \, \forall \, t$), we expect that 
\beq
\label{eq:lyap}
\langle \ln \delta q_j(t) \rangle_s = \lambda (t - t_j^{(0)}) + \ln \delta \, ,\quad t_j^{(0)} \leq t \leq t_j^* \, ,
\eeq
where $t_j^{(0)}$ denotes the time it takes the initial perturbation to reach the $j^{th}$ site and $t_j^*$ denotes the ($\delta$ dependent) saturation time. For a chaotic system, one expects that the propagation of the perturbation is ballistic and it will arrive at site $j$ at a time $t_j^{(0)} \sim j/v_b$, where $v_b$ is the butterfly velocity and the arrival time is defined as the time at which the deviation satisfies $\delta q_j(t_j^{(0)}) = \delta$~\cite{kumar2020}. Past the saturation time $t_j^*$ i.e., once $\delta q_j(t)$ saturates to its maximum attainable value, $\lambda_j(t) \sim 1/t$ and the finite-time Lyapunov exponent ceases to provide any physically meaningful information. 

In this paper, we study the experimentally relevant quantity $\lambda_N$, which we refer to as a \textit{non-local} Lyapunov exponent because it constitutes a non-local measurement of chaos in the system. Specifically, this quantity captures the response of the system to an infinitesimal perturbation in the coherent input, localized at one end of the chain, through the output signal radiating from the opposite end. We calculate $\lambda_N(t)$ for the distinct non-equilibrium steady states of the boundary-driven dissipative anharmonic chain and find that it provides a clear signature of chaos (or its absence) in the NESS. In particular, the behavior of $\lambda_N(t)$ exhibits two different regimes: (i) a regular, non-chaotic regime, where $\lambda_N(t) < 0 \, \forall \, t$ and (ii) a chaotic regime characterized by a positive exponent $\lambda_N > 0$ which we extract from the intermediate time behavior Eq.~\eqref{eq:lyap} since our system is bounded (see Sec.~\ref{sec:lyapunov}). Our results demonstrate that the non-local Lyapunov exponent $\lambda_N(t)$ provides a crisp, experimentally feasible measure for probing the presence (or absence) of chaos in boundary-driven dissipative nonlinear chains.

\paragraph*{Numerical procedure: } Our numerical procedure for obtaining the NESS is as follows: we numerically integrate Eqs.~\eqref{eq:maineq1}-\eqref{eq:maineqN} using a velocity Verlet algorithm that has been adapted to include the damping and driving~\cite{allen2017computer}. In order to obtain transport properties in the NESS, we first allow the system sufficient time to relax to the NESS before time-averaging. In the simulations, we integrate the dynamics with a time step $\Delta t= 0.005$ and allow a time $T = 10^7$ (or $2\times 10^9$ time steps) for the system to reach the NESS. Convergence to the NESS is checked through Eq.~\eqref{eq:NESScond}, with a flat energy current profile throughout the system signalling the stationary state. We then obtain time-averaged quantities in the NESS by simulating the equations of motion for an additional $2\times 10^8$ time steps, with measurements taken at every $2\times 10^4$ time-steps. 


\section{Analytic Solution for the Harmonic Chain}
\label{sec:linear}

We first consider the steady-state properties of a driven dissipative harmonic chain, for which the equations of motion are given by Eqs.~\eqref{eq:maineq1}-\eqref{eq:maineqN}, with $\nu = 0$. While the steady-state transport properties of a harmonic chain in contact with finite temperature reservoirs have been extensively studied (see e.g., Refs.~[\onlinecite{dhar2012,dhar2019,weiderpass2020}]), the boundary-driven dissipative case is less familiar. We first re-write the equations of motion compactly:
\beq
\label{eq:harmoniceom}
\ddot{\bm{q}} = -\bm{K} \bm{q} - \bm{D} \dot{\bm{q}} + \bm{f}(t) \, , 
\eeq
where $\bm{q} = (q_1,\cdots,q_N)^T$, $f_j(t) = \delta_{j,1} F_d \cos(\omega_d t)$, $D_{ij} = \eta \delta_{i,j} \left(\delta_{j,1} + \delta_{j,N}\right)$, and
\beq
\label{eq:Kdef}
K_{ij} = (1 + 2 \epsilon)\delta_{i,j} - \epsilon\delta_{i,j}(\delta_{j,1}+\delta_{j,N}) - \epsilon \delta_{|i-j|,1} \, .
\eeq

The steady-state solution to Eq.~\eqref{eq:harmoniceom} is straightforwardly found by going to the Fourier domain:
\begin{align}
\bm{q}(t) &= \int_{-\infty}^\infty d\omega  \, e^{-i\omega t} \bm{g}(\omega) \mathcal{F}\{\bm{f}\}(\omega) \nonumber \\
&= - \Im \left[\bm{g}(\omega_d) \bm{F} e^{-i \omega_d t} \right] 
\end{align}
where the Green's function $g(\omega)$ is given by
\beq
\bm{g}(\omega) = \left( \bm{K} - \omega^2\bm{I} - i \omega \bm{D}\right)^{-1} \, ,
\eeq
and $F_j = \delta_{j,1} F_d$. The steady-state solution is clearly periodic and is synchronized with the driving frequency $\omega_d$ for any system size $N$, as expected for a linear system. 

\begin{figure}[t]
    \centering
    \includegraphics[width=0.45\textwidth]{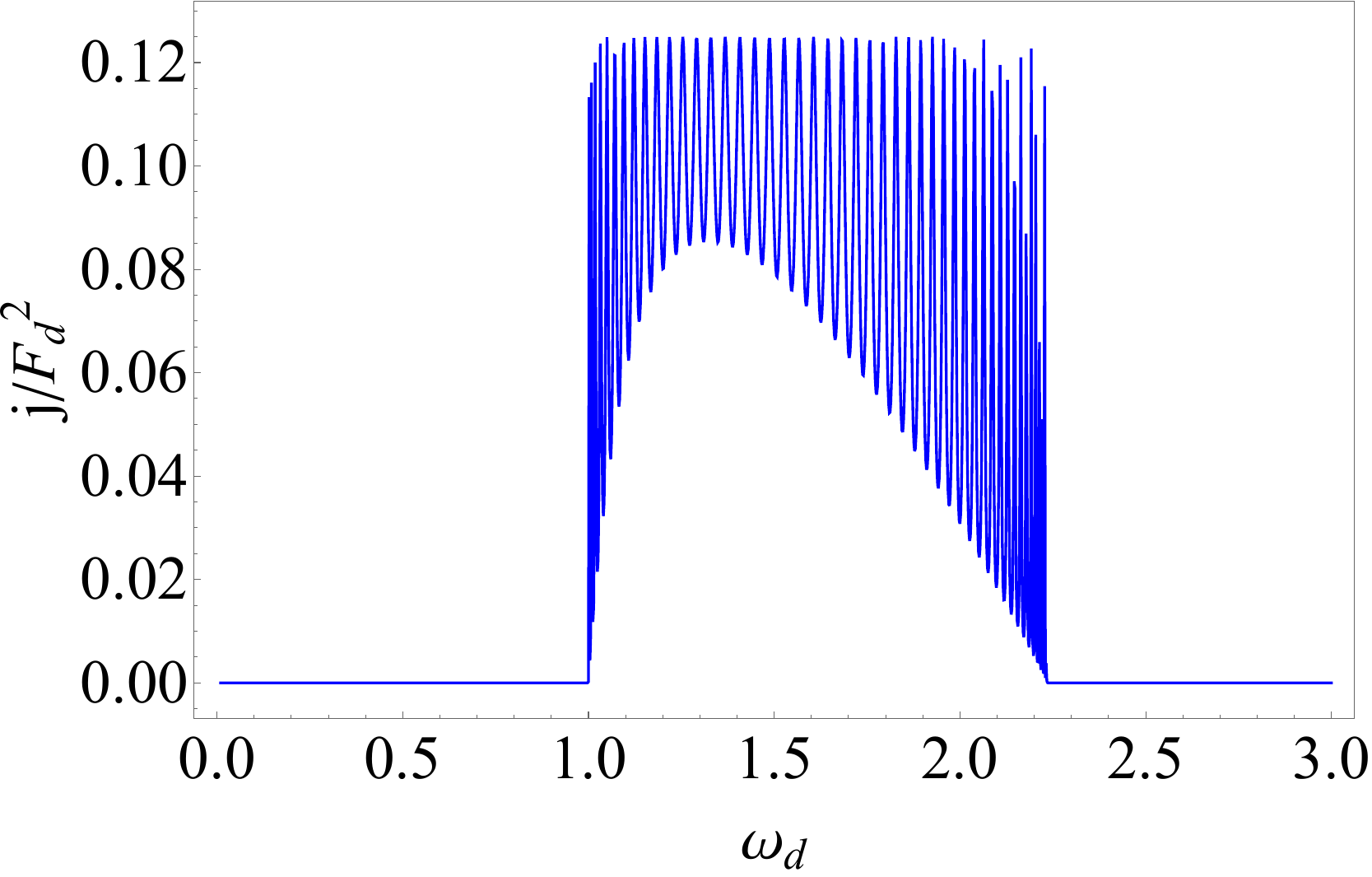}
    \caption{Steady-state energy current $j$ vs. driving frequency $\omega_d$ for the harmonic chain with $\epsilon = 1$, $\eta = 1$, $N = 50$. The energy current scales quadratically with the driving amplitude $F_d$.}
    \label{fig:harmonicJ}
\end{figure}

\begin{figure}[t]
     \centering
     \begin{subfigure}[b]{0.23\textwidth}
         \centering
         \includegraphics[width=\textwidth]{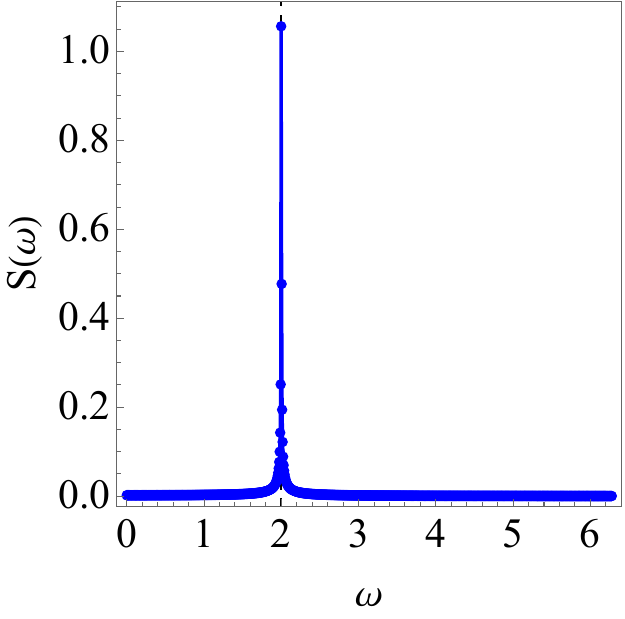}
         \caption{}
         \label{fig:harmonicSW}
     \end{subfigure}
     \hfill
     \begin{subfigure}[b]{0.23\textwidth}
         \centering
         \includegraphics[width=\textwidth]{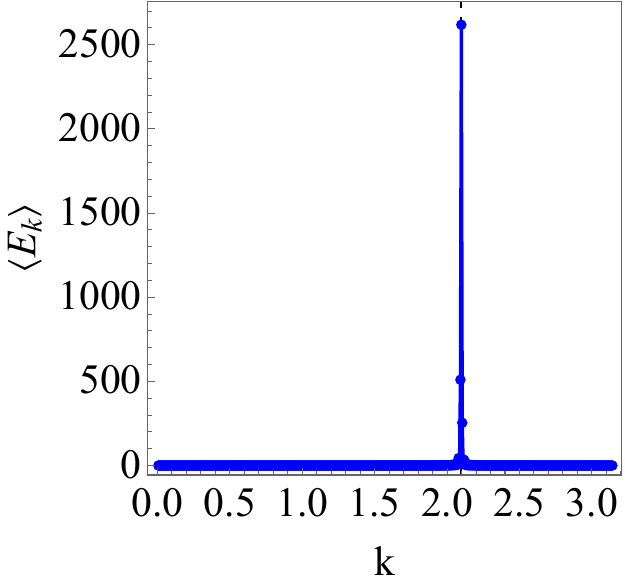}
         \caption{}
         \label{fig:harmonicEK}
     \end{subfigure}
         \\
     \begin{subfigure}[b]{0.23\textwidth}
         \centering
         \includegraphics[width=\textwidth]{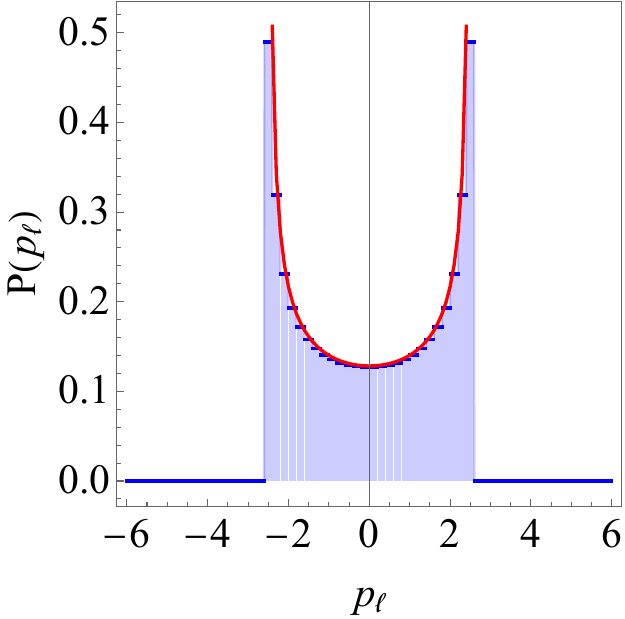}
         \caption{}
         \label{fig:harmonicdist}
     \end{subfigure}
     \hfill
     \begin{subfigure}[b]{0.23\textwidth}
         \centering
         \includegraphics[width=\textwidth]{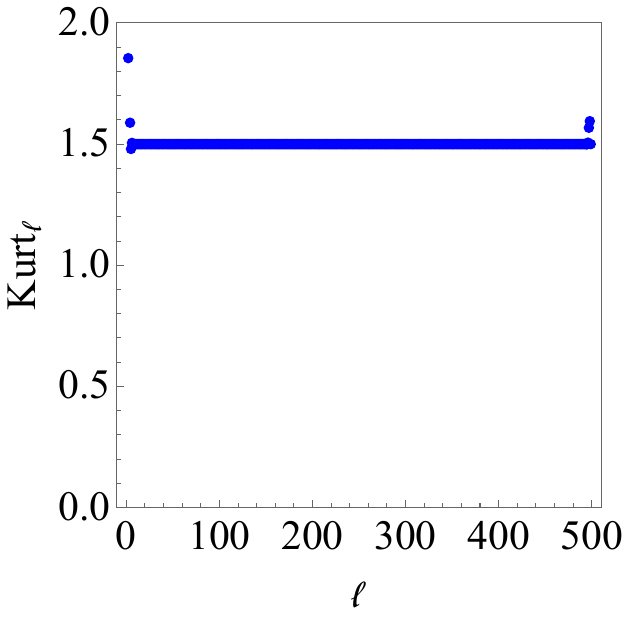}
         \caption{}
         \label{fig:harmonickurt}
     \end{subfigure} 
        \caption{Numerically obtained (a) power spectrum $S(\omega)$ and (b) time-averaged mode spectrum $\braket{E_k}$ in the steady-state of the boundary-driven dissipative harmonic chain with $\epsilon=1$, $\eta = 1$, $F_d = 10$, $\omega_d = 2$, and $N = 500$. The steady-state response is synchronized with the driving frequency (dashed black line in (a)) and contains only the resonant normal mode with quasimomentum $k^* \sim 2\pi/3$ (dashed black line in (b)) corresponding to $\omega_{k^*} \approx \omega_d$. (c) The local canonical momentum distribution (shown here for site $\ell = 245$) matches the analytic prediction Eq.~\eqref{eq:harmonicdist} (in red) as does (d) the kurtosis, which equals 1.5 in the bulk.}
        \label{fig:harmonicSWEK}
\end{figure}

From the above analytic solution, it is also straightforward to obtain the time-averaged steady-state energy currents\footnote{Here, we have used the fact that the steady-state solution is periodic, so that time-averaged quantities equal those averaged over a single driving period $\Omega$: 
\beq
\langle A \rangle = \lim_{T\to \infty} \frac{1}{T} \int_{T_0}^{T_0 + T} A(t) = \lim_{n \to \infty} \frac{\Omega}{2\pi n} \int_{0}^{\frac{2\pi n}{\Omega}} A(t) = \frac{2 \pi}{\Omega}\int_0^{\frac{2\pi}{\Omega}}A(t) .
\eeq
}
\begin{align}
    \langle j_{1,L} \rangle = &\frac{F_d^2 \omega_d}{2} \Re\left[g_{1,1}(\omega_d) \right] - \eta \frac{F_d^2 \omega_d^2}{2} |g_{1,1}(\omega_d)|^2 \nonumber \\
    \langle j_{\ell,\ell-1} \rangle =& \epsilon\frac{F_d^2 \omega_d}{2} \Im \left[ g_{\ell,1}(\omega_d) g^*_{\ell-1,1}(\omega_d) \right] \nonumber \\
    \langle j_{R,N} \rangle = &  \eta \frac{F_d^2 \omega_d^2}{2} |g_{N,1}(\omega_d)|^2 \, ,
\end{align}
for which one can show analytically that Eq.~\eqref{eq:NESScond} holds. For the harmonic chain, we thus find that the steady-state energy current scales quadratically with the driving amplitude $j \propto F_d^2$, independent of system size. In Fig.~\ref{fig:harmonicJ}, we plot the time-averaged steady-state energy current $j$ as a function of driving frequency $\omega_d$. As expected for the harmonic chain, there is no response for driving frequencies which lie outside the band.  For driving frequencies within the band, the observed multi-resonances result from the resonances that occur whenever the driving frequency approaches the eigenfrequencies of the harmonic chain, which take values in the range $\omega_k \in [1, \sqrt{1 + 4 \epsilon}]$\footnote{For strong (resp. weak) dissipation, one can show~\cite{zhang2011} that the resonant frequencies correspond to the eigenfrequencies of the matrix $\bm{K}$~\eqref{eq:Kdef} for system size $N-2$ (resp. $N$)}. We thus see that the steady-state solution for the boundary-driven dissipative harmonic chain corresponds to a periodic solution synchronized with the driving frequency $\omega_d$ and where only the normal mode with eigenfrequency $\omega_{k^*} \approx \omega_d$ survives in the long time limit. This is also confirmed numerically, as shown in Fig.~\ref{fig:harmonicSWEK} (note that $S(-\omega) = S(\omega)$).

The steady-state solution for the harmonic chain allows us to analytically compute the kurtosis~\eqref{eq:Kurtdef} as well: one can show that 
\beq
\langle p_\ell^2 \rangle = \frac{F_d^2 \omega_d^2}{2} |g_{\ell,1}(\omega_d)|^2\, , \quad \langle p_\ell^4 \rangle = \frac{3 F_d^4 \omega_d^4}{8}  |g_{\ell,1}(\omega_d)|^4 \, ,
\eeq
so that 
\beq
\text{Kurt}_\ell = 1.5 \quad \forall \, \ell \in [1,N]
\eeq
is uniform across the chain and is independent of the driving parameters as well as the system size. This value of the kurtosis reveals that the non-equilibrium steady state for the boundary-driven dissipative harmonic chain is non-thermal i.e., local observables are not distributed according to the Maxwell-Boltzmann distribution. In fact, we can analytically extract the probability distribution for the momentum from the exact steady-state:
\beq
\label{eq:harmonicdist}
P(p_\ell) = \frac{1}{\pi \left(F_d^2 \omega_d^2 |g_{\ell,1}|^2 - p_\ell^2 \right)^{1/2}} \, ,
\eeq
which corresponds to the bimodal arcsine distribution. As shown in Figs.~\ref{fig:harmonicdist} and~\ref{fig:harmonickurt}, there is excellent agreement between the numerically obtained steady-state and predictions for the exact steady-state.

For a harmonic chain with boundary couplings to thermal reservoirs at different temperatures, the steady-state reaches thermal equilibrium with a temperature that is uniform in the bulk and the model exhibits ballistic energy transport i.e., the energy current $j$ is independent of system size, in violation of Fourier's law~\cite{rieder1967,dhar2018}. Similarly, we find that the energy current in the boundary-driven dissipative harmonic chain is independent of system size: $j \sim N^0$, consistent with ballistic transport. However, the NESS in this case (for driving frequencies within the harmonic band) is a non-ergodic, non-thermal energy-transporting steady-state that is driven far from equilibrium and, in the thermodynamic limit, corresponds to a single resonant normal mode synchronized with the driving frequency. 


\section{Dynamical Regimes of the Discrete Klein-Gordon Chain}
\label{sec:kg}
We now turn to the more interesting case of the discrete nonlinear Klein-Gordon chain whose Hamiltonian is given by Eq.~\eqref{eq:HSys} with $\nu = 1$. Previous works have studied the onset of thermalization and equipartition in the closed Klein-Gordon Hamiltonian as a consequence of the nonlinear interactions between normal modes of the discrete lattice~\cite{pistone2018,danieli2019}. When this nonlinear system is coupled at its two ends to thermal reservoirs at different temperatures, it relaxes\footnote{For such anharmonic chains, convergence to the NESS, when it can be established, only follows a stretched exponential due to the presence of breathers~\cite{tsironis96,hairer2009}.} to a current-carrying thermal NESS with an average steady-state current that scales with system size as $\sim 1/N$, in accordance with Fourier's law~\cite{aoki2005,lefevere2006,li2007}. Transport in this system is generically expected to be diffusive and, in the limit of low bath temperatures, the temperature profile in the NESS displays a linear gradient $\sim 1/T$ across the bulk of the chain~\cite{dhar2008}.

While the effect of coupling discrete nonlinear chains to incoherent thermal reservoirs has been extensively investigated, periodically driven nonlinear chains---where energy is injected coherently into the system---have been less explored. Studies in this direction typically consider the limit of weak anharmonicity and bulk driving/dissipation, where translation invariance is restored~\cite{burlakov1998,vanossi2000,marin2001,khomeriki2001,maniadis2006,hennig2008}. Here, we instead study the boundary-driven dissipative discrete nonlinear Klein-Gordon chain, where the periodic drive explicitly breaks translation invariance, and do not restrict ourselves to the limit of weak driving (equivalently, weak anharmonicity). The equations of motion are given by Eqs.~\eqref{eq:maineq1}-\eqref{eq:maineqN} with $\nu = 1$. In what follows, we set $\epsilon = 1$, $\eta = 1$, and focus on the NESS resulting from in-band driving i.e., on frequencies $\omega_d$ that lie within the harmonic band $\omega_k \in [1,\sqrt{1 + 4 \epsilon]}$.

\begin{figure}[t]
    \centering
    \includegraphics[width=0.45\textwidth]{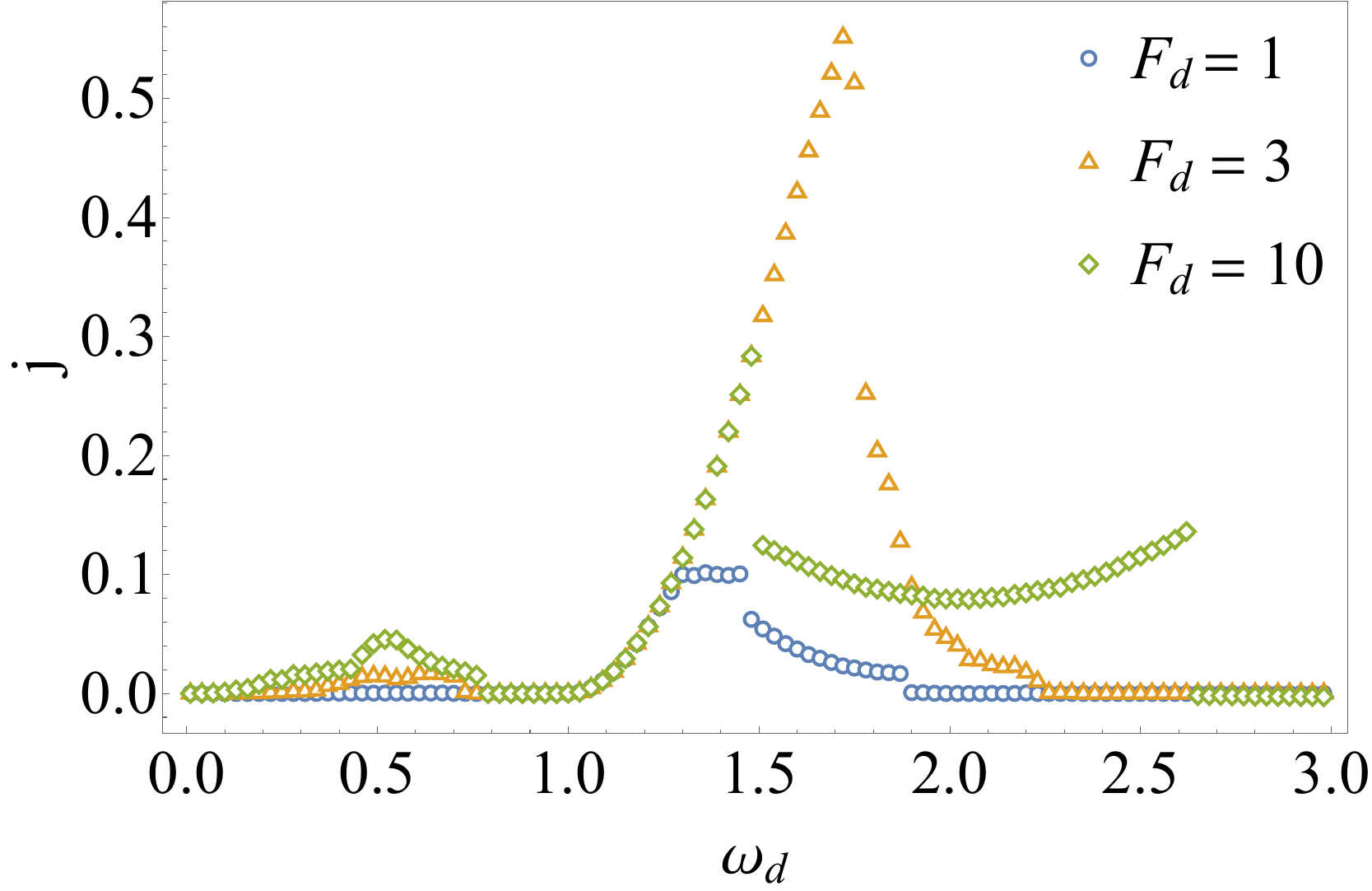}
    \caption{Numerically obtained steady-state energy current $j$ as a function of driving frequency $\omega_d$ for the anharmonic chain with $N = 500$ and three distinct driving amplitudes.}
    \label{fig:anharmonicJ}
\end{figure}

The steady-state frequency response curve for the system is shown in Fig.~\ref{fig:anharmonicJ} for three different driving amplitudes $F_d$, where we see that the nonlinearity smooths out the multi-resonant behavior observed in the harmonic case. In marked contrast with the the harmonic case, we observe a non-vanishing energy current at driving frequencies which lie outside the harmonic band, a phenomenon referred to as ``nonlinear supratransmission"~\cite{geniet2002,khomeriki2004}. In particular, besides a non-vanishing response at driving frequencies below the lower band edge ($\omega_k = 1$), we find that the anharmonicity also effectively extends the upper band edge up to a value that depends nonlinearly on the driving amplitude. While we do not have an analytic expression for the effective upper band edge, we numerically observe that it increases monotonically with the driving amplitude.

\begin{figure}[t]
     \centering
     \begin{subfigure}[b]{0.45\textwidth}
         \centering
         \includegraphics[width=\textwidth]{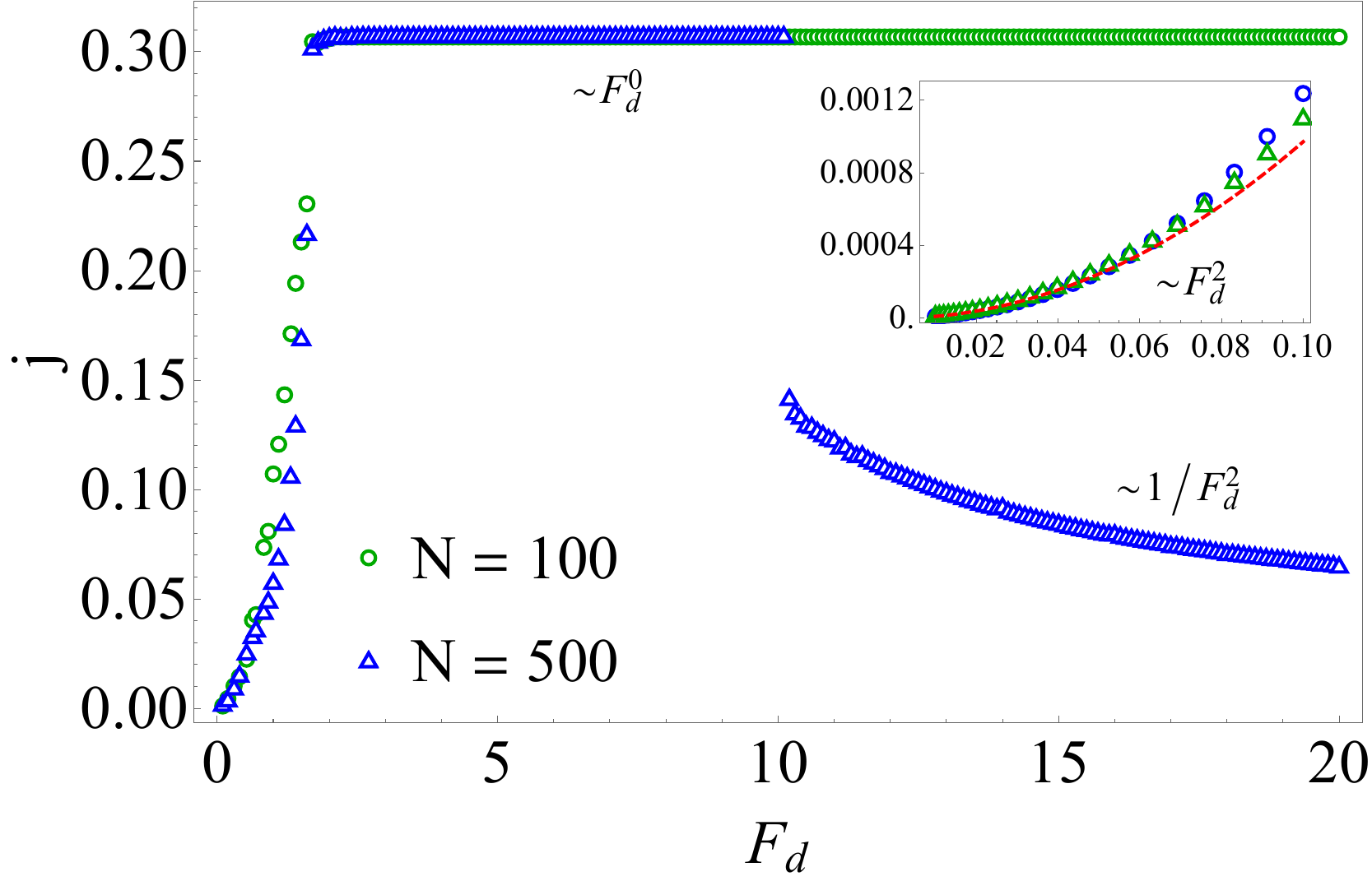}
         \caption{}
         \label{fig:JvsFmain}
     \end{subfigure}
     \\
     \begin{subfigure}[b]{0.45\textwidth}
         \centering
         \includegraphics[width=\textwidth]{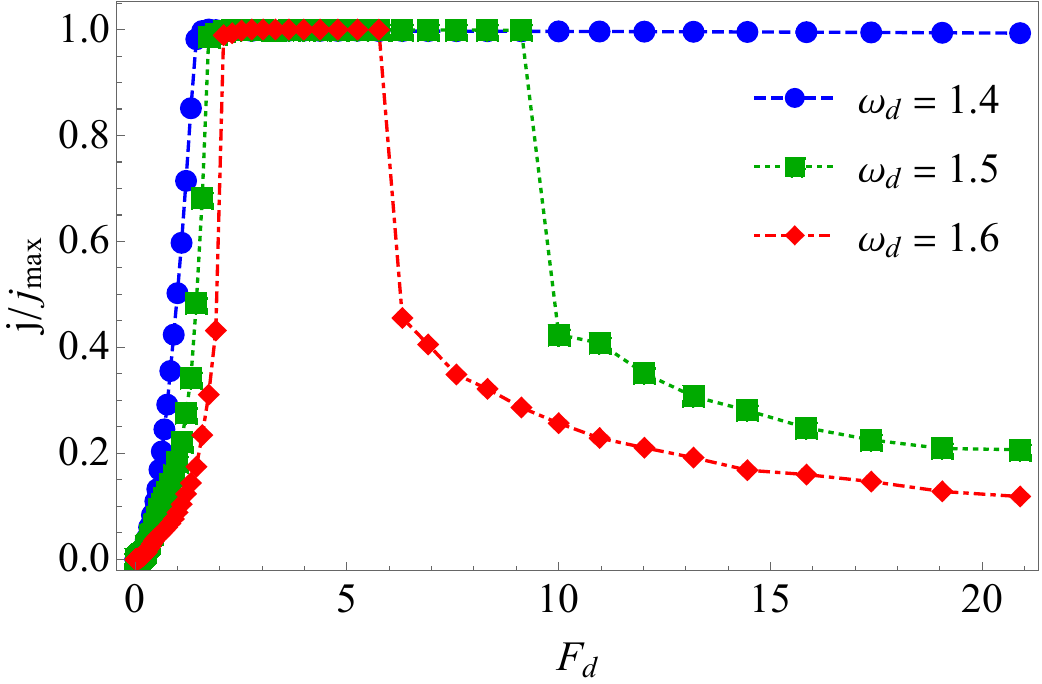}
         \caption{}
         \label{fig:JvsFmain_2}
     \end{subfigure}
        \caption{(a) Numerically obtained steady-state energy current $j$ as a function of driving amplitude $F_d$ for the anharmonic chain with $\omega_d = 1.5$ and $N = 100, 500$. The inset shows small driving amplitudes and the dashed red line indicates the expected behavior within the single nonlinear phonon approximation (see Sec.~\ref{sec:weak}).(b) For fixed system size ($N = 500$ here), the width of the intervening plateau region (where $j \sim F_d^0$) is sensitive to the driving frequency. To enable a direct comparison, the current is normalized by its maximum value as a function of $F_d$.}
        \label{fig:JvsF}
\end{figure}

From Fig.~\ref{fig:anharmonicJ}, it is also evident that the energy current no longer behaves simply as $j \sim F_d^2$; instead, the current \textit{decreases} when the driving amplitude is increased from $F_d = 3$ to $F_d = 10$. For driving frequencies within the harmonic band, the characteristic behavior of the energy current as a function of the driving amplitude is illustrated in Figs.~\ref{fig:JvsFmain} and~\ref{fig:JvsFmain_2}, which show four qualitatively distinct dynamical steady-state regimes:
\begin{itemize}
    \item[(A)]  For low driving amplitudes (equivalently, weak nonlinearity), the energy current scales as $j \sim F_d^2$, which corresponds to the weakly perturbed quasilinear regime where a single nonlinear phonon is resonantly driven. The dynamical properties of this non-ergodic regime with ballistic transport are discussed in Sec.~\ref{sec:weak}. 
    \item[(B)] As shown in the inset of Fig.~\ref{fig:JvsFmain}, the quasilinear approximation breaks down at some driving strength and the current displays a steeper dependence on the driving amplitude: $j \sim F_d^\alpha$, with $\alpha > 2$. This increase indicates that higher order nonlinear effects must be accounted for and lead to multiphonon resonances. See Sec.~\ref{sec:thermal} for details on this regime, which is in local thermal equilibrium and displays super-diffusive energy transport. For all parameters we have studied, the crossover from region (A) to (B) occurs smoothly. 
    \item[(C)] Above a certain threshold $F_{c1}$, the current saturates to a constant value that is independent of $F_d$ and $N$, but is a function of $\omega_d$. This corresponds to the \textit{resonant nonlinear wave} regime, which supports ballistic transport and which we discuss in Sec.~\ref{sec:DDBC}. 
    \item[(D)] Above a second threshold $F_{c2}$, the resonant nonlinear wave is destabilized and the energy current abruptly drops, decaying as $j \sim 1/F_d^2$. As we discuss in Sec.~\ref{sec:mixed}, this regime of low transmission corresponds to an unusual steady-state with diffusive transport, in which only a certain spatial region of the chain is in local thermal equilibrium, while the rest is \emph{not} in local thermal equilibrium.
\end{itemize}
Similar to what we find here, a transition between a non-ergodic region (A) where the current grows with the driving strength to a chaotic steady-state (D) where the current decreases with stronger driving has been reported numerically~\cite{debnath2017} and observed experimentally~\cite{fitzpatrick2017,fedorov2021} in driven-dissipative circuit QED arrays working in the semiclassical regime of large photon number. However, to our knowledge, the intermediate regimes (B and C) that we find here have not previously been discussed in discrete driven-dissipative classical or quantum systems\footnote{We remark though that the appearance of a linearly stable spatial pattern preceding the onset of chaos in the boundary-driven dissipative anharmonic chain is similar to the bulk-driven dissipative FPUT lattice~\cite{khomeriki2001}, where stable patterns appear before the transition to chaos.}.

Before elaborating on the properties of the dynamical regimes described above, we briefly discuss the dependence of the thresholds $F_{c1}$ and $F_{c2}$ on the driving frequency $\omega_d$ and system size $N$. For in-band driving, we observe only a weak dependence of the first threshold $F_{c1}$ on $\omega_d$ and $N$ (see Fig.~\ref{fig:JvsF}). However, the second threshold $F_{c2}$ is strongly dependent on $\omega_d$ and $N$. As illustrated in Fig.~\ref{fig:JvsFmain_2}, for fixed system size, $F_{c2}$ monotonically decreases as $\omega_d$ is tuned from the lower band edge to the upper band edge of the harmonic lattice; for driving frequencies above the upper harmonic band edge, there is no intermediate region (C) for any system size and the dynamics transitions directly from (B) to (D) at a critical driving amplitude that depends on the driving frequency $\omega_d$ and system size $N$. Since we are only considering in-band driving, this direct transition is beyond the scope of this work. 

\begin{figure}[t]
    \centering
    \includegraphics[width=0.4\textwidth]{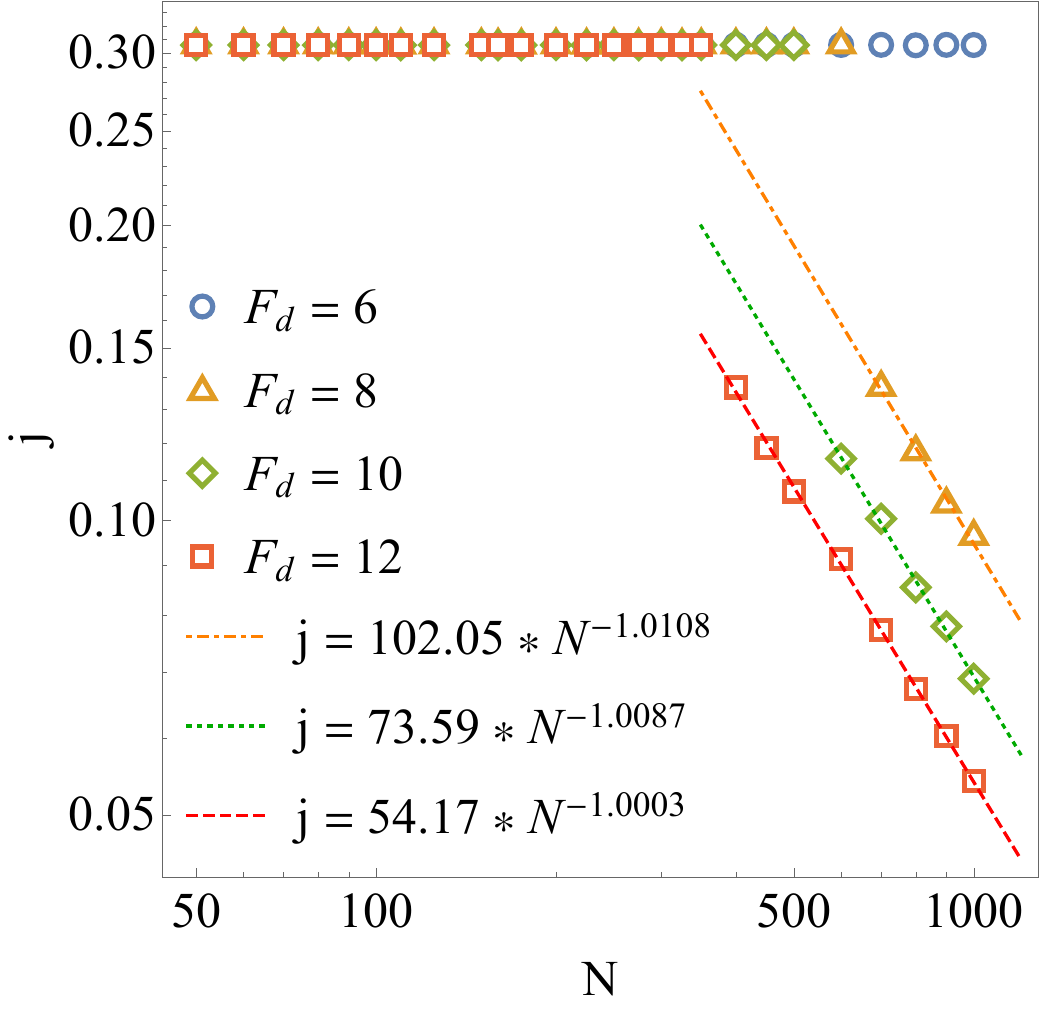}
    \caption{Scaling of the steady-state energy current $j$ with system size $N$ for different values of driving amplitude $F_d$ at fixed driving frequency $\omega_d = 1.5$. Markers denote numerical data points and the dashed red line indicates the corresponding fit.}
    \label{fig:Scaling2}
\end{figure}

Fig.~\ref{fig:JvsFmain} also illustrates the system-size dependence of the second threshold $F_{c2}$, which we find decreases with increasing $N$. We observe a power-law relation but since the exponent is sensitive to the driving frequency $\omega_d$ as well, we are unable to find a quantitative characterization. Based on this observation, we expect that the intervening region (C) will be absent in the thermodynamic limit, where the dynamics will transition directly from (B) to (D).

We further characterize the system-size dependence in Fig.~\ref{fig:Scaling2}, which displays the NESS averaged steady-state current $j$ as a function of $N$ for various large driving amplitudes $F_d$. For short chains $j \sim N^0$, which indicates ballistic transport and (for $F_d > F_{c1}$) corresponds to the resonant nonlinear wave regime (C). However, above a certain chain length $N_c$ the energy current abruptly drops and decays as $j \sim N^{-1}$, indicating diffusive transport and an onset of chaos. In Sec.~\ref{sec:DDBC}, we argue that the resonant nonlinear wave is linearly stable, which shows that this crossover from ballistic to diffusive transport is a consequence of a bulk nonlinear instability, which is not captured by linear response. Fig.~\ref{fig:Scaling2} also shows that $N_c$ depends on the driving amplitude $F_d$, such that stronger driving induces this instability at smaller system sizes, consistent with the fact that stronger driving corresponds to stronger nonlinearity in our model. We also find that $N_c$ decreases as the driving frequency is increased from the lower harmonic band edge to the upper harmonic band edge. However, even more extensive numerical simulations are required to find the exact relation between $N_c$ and the driving parameters $(\omega_d, F_d)$.

Based on the above observations, we conclude that the intermediate dynamical regime (C) is absent in the thermodynamic limit, where transport will generally be diffusive (or possibly super-diffusive~\cite{Spohn_2014}) away from the weakly perturbed quasilinear regime, which shows ballistic transport. However, since circuit QED experiments typically operate at a fixed system size and study the steady-state dynamics as a function of the external driving parameters, our results suggest that the resonant nonlinear wave will appear as a distinct dynamical regime at system sizes accessible to current experiments.

We now discuss each of the regimes (A)-(D) and the distinct dynamical features of their current-carrying NESS in detail, followed by results on the non-local Lyapunov exponent.


\subsection{Weakly Perturbed Quasilinear Regime}
\label{sec:weak}

\begin{figure}[t]
    \centering
    \includegraphics[width=0.5\textwidth]{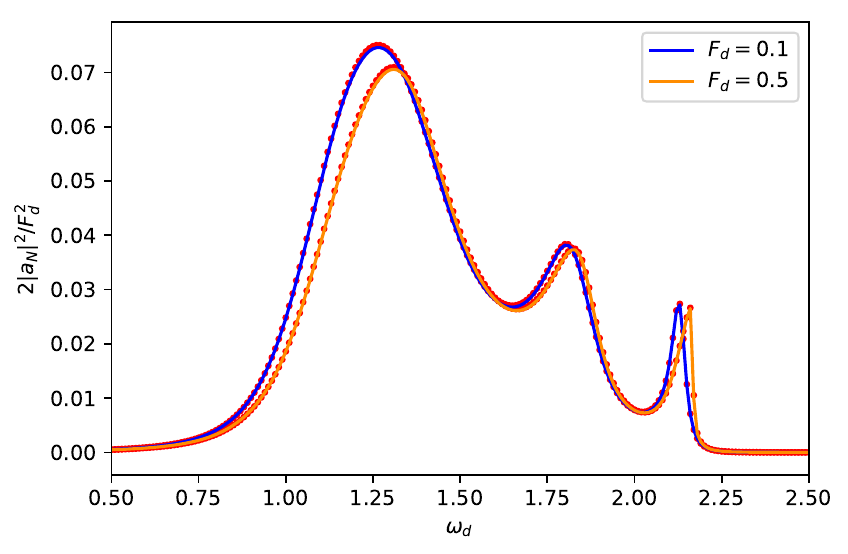}
    \caption{For an anharmonic chain of $N=5$ oscillators, the perturbative harmonic balance condition Eq. \eqref{eq:foldover} (solid lines) accurately captures the numerically observed nonlinear response of the boundary-driven dissipative Klein-Gordon chain (red markers) up to the onset of multi-valued response at $F_d \approx 0.5$.}
    \label{fig:foldover}
\end{figure}

In the perturbative regime of weak driving (equivalently, weak nonlinearity), the boundary-driven dissipative Klein-Gordon chain exhibits a many-body version of the ``foldover'' effect familiar from the theory of nonlinear resonance in few-body anharmonic oscillators, such as the Duffing oscillator. We can capture this approximately through the harmonic-balance approximation, applied to the nonlinear chain Eq. \eqref{eq:maineqN}, which yields the system 
\begin{equation}
\left(-\omega_d^2\delta_{ij} + i\omega_d D_{ij} + K_{ij}+3 |a_i|^2\delta_{ij}\right)a_j = (F_d/2)\delta_{i,1}
\label{eq:foldover}
\end{equation}
of $N$ coupled nonlinear equations for the steady-state response $q_j = a_je^{i\omega_dt}+a_j^*e^{-i\omega_dt}$ at the driving frequency. For sufficiently weak driving, which corresponds to weak nonlinearity (see Eq.~\eqref{eq:relabel}), the onsite positive-frequency response $a_j(\omega_d)$ at the driving frequency is single-valued and the harmonic-balance equation Eq. \eqref{eq:foldover} can be solved by numerical iteration, yielding foldover of each single-phonon resonance that is qualitatively similar to what is seen in the Duffing oscillator and quantitatively matches direct numerical simulations of the anharmonic chain, as depicted in Fig.~\ref{fig:foldover}. Beyond the perturbative regime, the frequency response predicted by Eq. \eqref{eq:foldover} becomes multi-valued, and the iterative solution no longer converges.

Nevertheless, we can further characterize this regime by looking for solutions $q_j(t) = F_d\cos(k j -\omega_d t)$, which correspond to extended quasiharmonic waves i.e., nonlinear phonons. Within the rotating wave approximation and ignoring the boundary dissipation, we can approximate the nonlinear dispersion relation as~\cite{khomeriki2004}
\beq
\tilde{\omega}_k(F_d) = \sqrt{1 + 4\epsilon \sin^2\left(\frac{k}{2}\right)+ \frac{3}{4} F_d^2} \, ,
\eeq
with quasimomentum $k \in [0,\pi)$. If we make the single nonlinear phonon approximation, within which only those resonant phonons are excited whose quasimomenta satisfy $\tilde{\omega}_k(F_d) = \omega_d$, we find that the energy current $j \propto F_d^2$. For very small driving amplitudes, this simple analysis is in agreement with our numerical results (see inset of Fig.~\ref{fig:JvsFmain}). In the weakly perturbed quasilinear regime, the NESS thus corresponds to a single nonlinear phonon resonant with the driving frequency. Consequently, the dynamics in this regime are qualitatively indistinguishable from the regular, non-chaotic dynamics of the harmonic system, with a negative Lyapunov exponent and a uniform kurtosis Kurt$_\ell = 1.5$ that is consistent with a non-thermal NESS.


\subsection{Thermal Regime with Super-diffusive Transport}
\label{sec:thermal}

\begin{figure}
    \centering
    \includegraphics[width=0.5\textwidth]{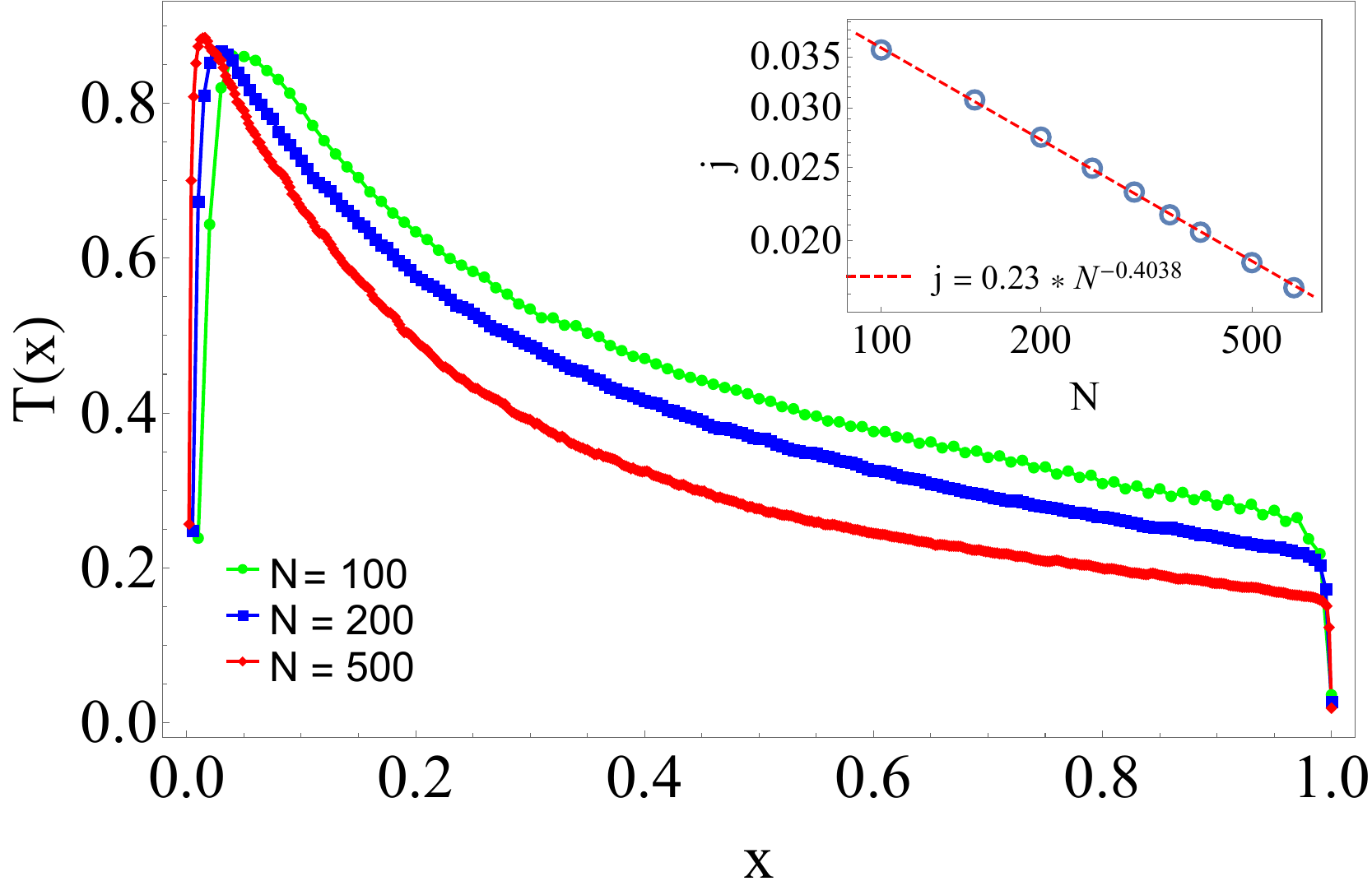}
    \caption{Temperature profiles for the anharmonic chain with $\omega_d=1.8$, $F_d = 1$. The temperature profile clearly displays a gradient across the bulk of the system. Inset: the steady-state energy current decays with system size as $\sim N^{-0.40}$, indicating that transport is super-diffusive in this regime. Markers denote numerical data points and the dashed red line indicates the corresponding fit.}
    \label{fig:temp1}
\end{figure}

\begin{figure}[t]
     \centering
     \begin{subfigure}[b]{0.23\textwidth}
         \centering
         \includegraphics[width=\textwidth]{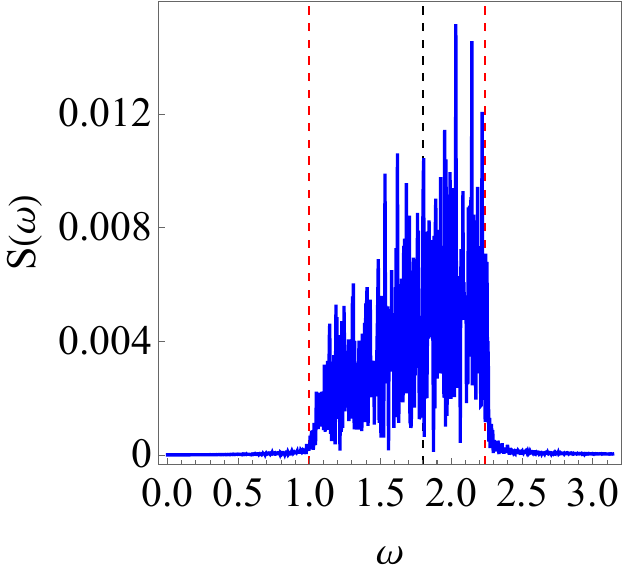}
         \caption{}
         \label{fig:anharmonicSW1}
     \end{subfigure}
     \hfill
     \begin{subfigure}[b]{0.23\textwidth}
         \centering
         \includegraphics[width=\textwidth]{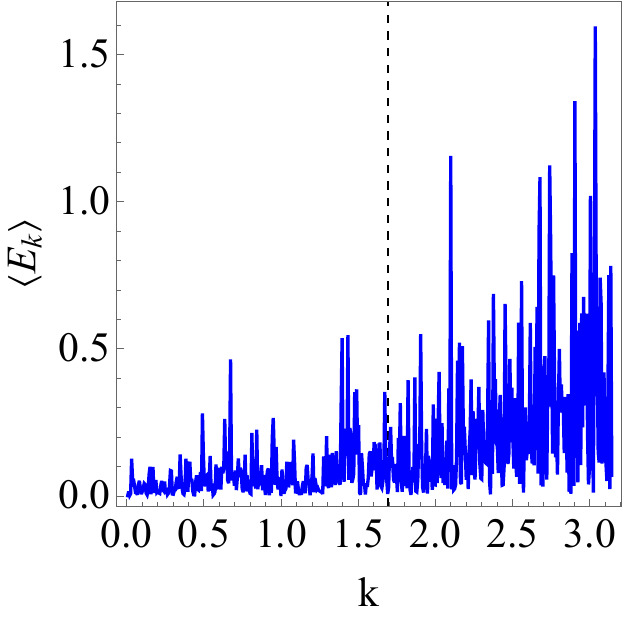}
         \caption{}
         \label{fig:anharmonicEK1}
     \end{subfigure}
         \\
     \begin{subfigure}[b]{0.23\textwidth}
         \centering
         \includegraphics[width=\textwidth]{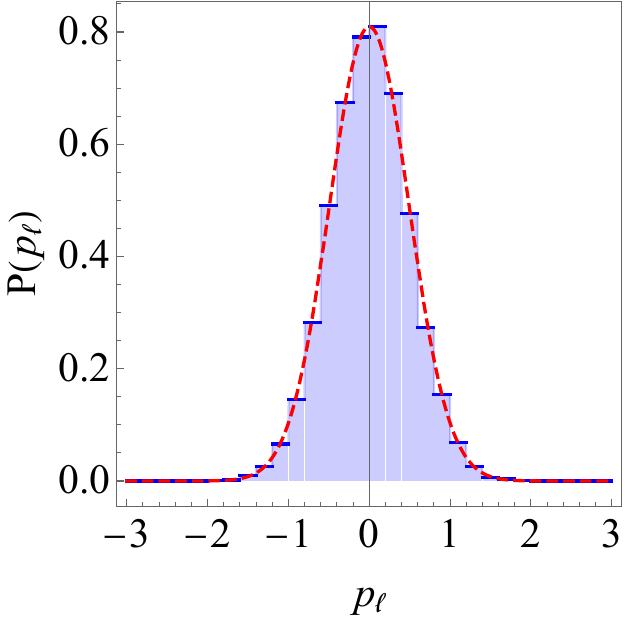}
         \caption{}
         \label{fig:anharmonicdist1}
     \end{subfigure}
     \hfill
     \begin{subfigure}[b]{0.23\textwidth}
         \centering
         \includegraphics[width=\textwidth]{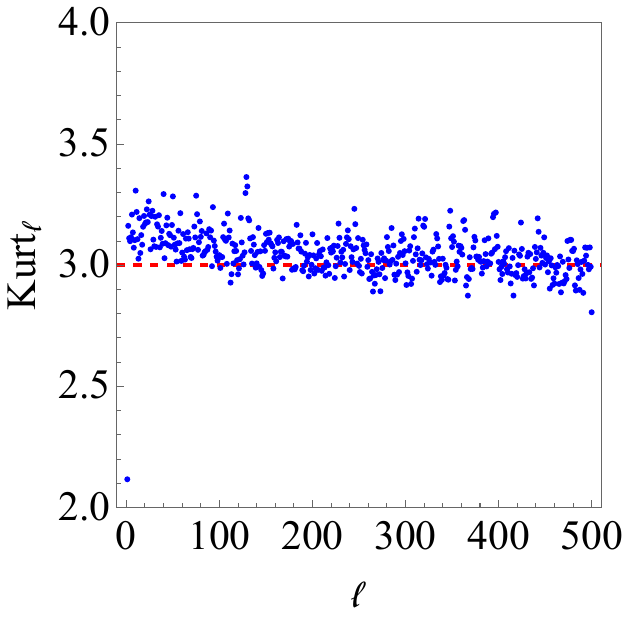}
         \caption{}
         \label{fig:anharmonickurt1}
     \end{subfigure} 
        \caption{Numerically obtained (a) power spectrum $S(\omega)$ and (b) time-averaged mode spectrum $\braket{E_k}$ in the steady-state of the boundary-driven dissipative anharmonic chain with $F_d = 1$, $\omega_d = 1.8$, and $N = 500$. The ergodicity of the NESS in this regime is reflected in the broadband emission spectrum and multi-phonon resonances. Dashed red lines in (a) indicate the harmonic-band edges, while dashed black lines in (a), (b) indicate the driving frequency $\omega_d$ and the quasimomentum corresponding to $\omega_{k^*} \approx \omega_d$. (c) The local canonical momentum distribution (shown here for site $\ell = 305$) matches the Maxwell-Boltzmann distribution Eq.~\eqref{eq:MBdist} (in red) with temperature set by the local temperature $T_\ell$. (d) Away from the externally driven left boundary, the kurtosis shows good agreement with the thermal expectation value (red dashed line).}
        \label{fig:spec1}
\end{figure}

On increasing the driving strength, the quasilinear approximation (which describes a single resonant quasimomentum synchronized with the drive) breaks down and, as shown in Fig.~\ref{fig:JvsFmain}, the observed energy current displays a steeper increase with $F_d$. This is consistent with the expectation that increasing the driving strength, which corresponds to increasing the nonlinearity, should lead to multiphoton resonances and an increase in the steady-state averaged energy current. For driving frequencies $\omega_d$ within the harmonic band and for driving strengths $0.5 \leq F_d < F_{c1}$, we find steady states with qualitatively similar features\footnote{For $F_d < 0.5$ but large enough that the quasilinear approximation is invalid, we do not find uniform behavior of the NESS. We leave a detailed investigation of this crossover from ballistic to super-diffusive transport to future work.}; we thus focus on these regions of parameter space in what follows and show that they correspond to the thermal regime of this model.

We show the NESS averaged local temperature profile for different chain lengths in Fig.~\ref{fig:temp1} for a representative set of driving parameters corresponding to this regime ($\omega_d = 1.8, F_d = 1.0$). Away from the boundaries, the NESS displays a non-trivial temperature gradient across the chain which resembles that of steady-states of anharmonic chains driven by thermal reservoirs at different temperatures~\cite{dhar2008}. We note that a similar temperature profile was also observed in the discrete nonlinear Schr\"odinger equation subject to a constant boundary forcing~\cite{iubini2014}. The inset in Fig.~\ref{fig:temp1} shows that the steady-state averaged energy current $j$ scales with chain length $N$ as $j \sim N^{-0.40}$, which indicates anomalous \textit{super-diffusive} transport, in violation of Fourier's law. We observe only a weak dependence of the exponent on the driving parameters for NESS within this regime. 

We probe this super-diffusive regime further by computing its power spectrum $S(\omega)$ and time-averaged mode spectrum $\braket{E_k}$ in the NESS, which are displayed in Fig.~\ref{fig:spec1}. In contrast with the synchronized response of the quasilinear regime, the emission spectrum here shows a broadband response that is non-trivial for all frequencies that lie within the harmonic band (indicated by dashed red lines) and vanishes otherwise. Similarly, $\braket{E_k}$ displays multiphoton resonances as well as ergodicity in the space of normal modes. This is expected since the boundary drive breaks translation symmetry and drives all normal modes of the harmonic lattice (albeit with varying strength). Along with the broadband emission spectrum, this reveals a chaotic current-carrying non-equilibrium steady state.

Surprisingly, we find that the local canonical momentum distributions in the bulk are Maxwell-Boltzmann distributed, with a temperature set by the local temperature $T_\ell$ at site $\ell$. This is shown in Fig.~\ref{fig:anharmonicdist1}, where the dashed red line corresponds to Eq.~\ref{eq:MBdist} with $T = T_\ell$ and no fitting parameter is involved. For sites away from the left boundary, which is being driven, the kurtosis is narrowly distributed around the thermal expectation value Kurt$_\ell = 3$, which demonstrates that the stationary state is locally equilibrated with respect to the thermal Maxwell-Boltzmann distribution. Hence, as the driving strength is increased away from the quasilinear regime, the anharmonic chain enters a regime that supports super-diffusive transport and a locally thermal, current-carrying NESS. The fact that the system is locally equilibrated indicates that the observed superdiffusion is a genuine hydrodynamic effect, that arises from strong interactions between normal modes rather than proximity to the harmonic chain.

We leave a detailed discussion of the appropriate hydrodynamic model for this anomalous heat conduction to future studies, and restrict ourselves here to some speculative remarks on the origin of this phenomenon. We first note that at non-zero temperature, the bulk model is expected to have no local conserved charges beyond the energy density. The na{\"i}ve expectation for single-mode hydrodynamics in one dimension is normal diffusion, and indeed this is the conclusion of a perturbative calculation in the classical Klein-Gordon chain~\cite{aoki2005}. The latter calculation predicts a thermal conductivity $\kappa \propto 1/T^2$ at low temperatures, which implies nonlinear diffusion of heat. A weaker divergence as $T \to 0$ would be consistent with superdiffusion of heat, and precisely this scenario has been argued to arise in one-dimensional metals~\cite{bulchandani2020superdiffusive}. However, the predicted $1/T^2$ divergence of the thermal conductivity, combined with the expectation that the bulk energy density recovers the classical phonon gas result $h \propto T$ at low temperatures, implies that the nonlinear diffusion model ceases to be well-posed as $T \to 0$ (see Ref.~\cite{bulchandani2020superdiffusive} and references therein). This suggests that a different theoretical approach is needed, since the rightmost edge of our system is always at zero temperature.

Another possibility is that the system has additional slow modes in addition to the energy density. This scenario is suggested by the fact that the super-diffusive regime succumbs at stronger driving to a dynamical regime with approximately conserved $U(1)$ phase differences (see Sec.~\ref{sec:DDBC} below), which might lead to hydrodynamic momentum-like modes associated with this conservation law. In the discrete nonlinear Schr{\"o}dinger chain at low temperature, such modes are known to give rise to super-diffusive energy transport~\cite{Mendl_2015}, and it is possible that a similar mechanism is at work here. In such cases, energy transport can be modelled as a L{\'e}vy walk, leading to the prediction that steady-state temperature profiles are described by the fractional diffusion equation~\cite{dhar2019anomalous}. 

Distinguishing conclusively between the two scenarios outlined above requires further simulations beyond the scope of this work; one diagnostic would be the scaling of the effective finite-size thermal conductivity $\kappa(N,T)$, which will increase with the system size $N$ if additional slow modes are responsible for the observed phenomenon.


\subsection{Resonant Nonlinear Wave Regime}
\label{sec:DDBC}

As the nonlinearity and driving strength are increased away from the weak-coupling limit, the na{\"i}ve expectation in a generic system is that the dynamics becomes chaotic, yielding a broadband frequency response and good mode-sharing between the normal modes of the linear spectrum. For the boundary-driven dissipative Klein-Gordon chain, these expectations are met as the drive amplitude is increased past the weak-coupling limit, as discussed in the previous section. Similarly, at sufficiently high driving amplitudes, the steady-state dynamics is also ergodic, which we discuss in the next section. Intriguingly however, for finite system sizes at intermediate driving amplitudes $F_{c1} < F_d < F_{c2}$, we find that this model supports a ``resonant nonlinear wave" regime, which corresponds to the plateau region in Fig.~\ref{fig:JvsFmain}. This far-from-equilibrium steady-state is characterized by several unusual properties, including a non-thermal NESS with a frequency response that is predominantly peaked about the driving frequency, emergent translation invariance, non-ergodicity in the space of Fourier modes, and ballistic transport. We note that while our numerics suggest a sharp transition from (B) to (C), a more detailed analysis is required to verify if this is indeed the case.
\begin{figure}
    \centering
    \includegraphics[width=0.5\textwidth]{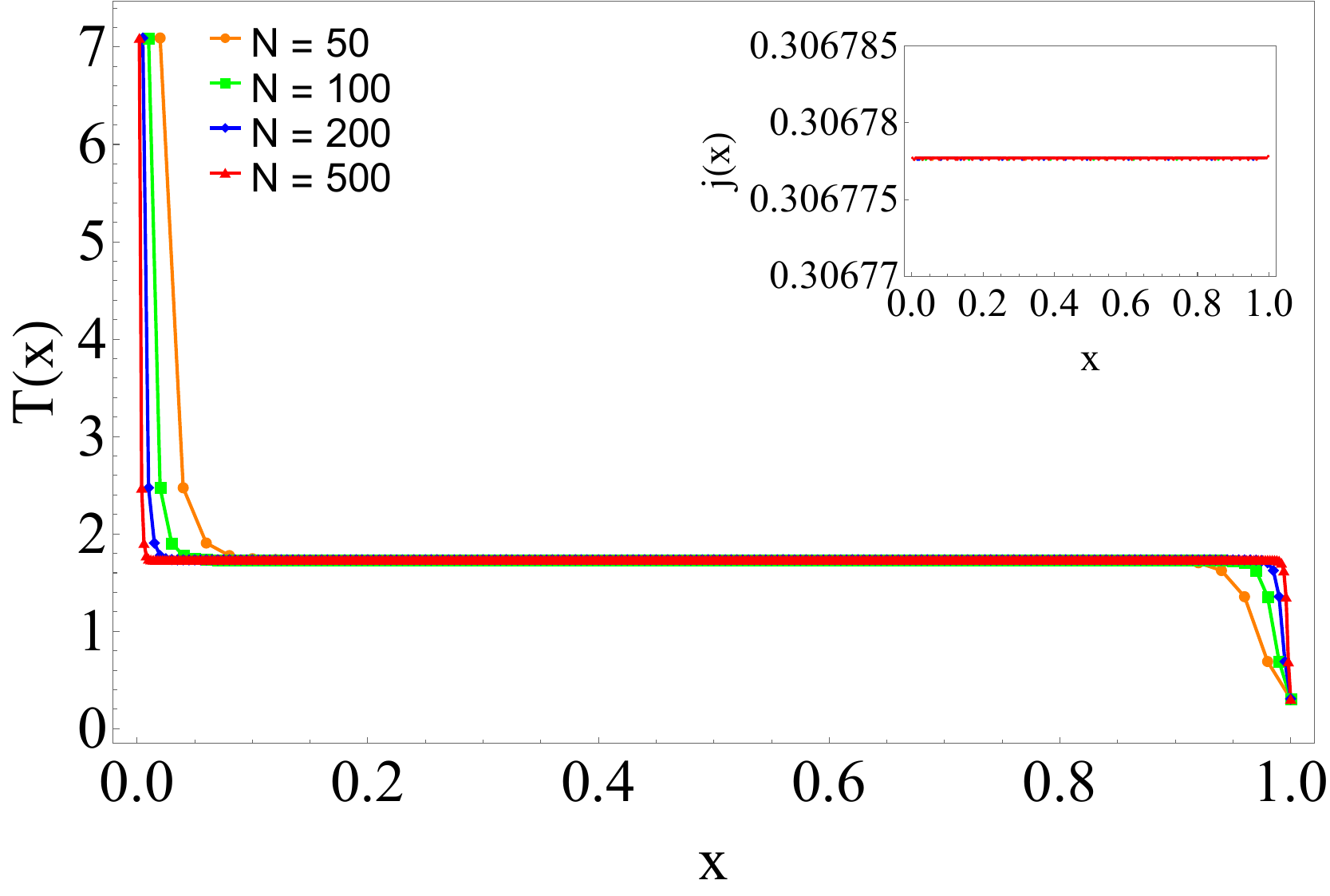}
    \caption{Temperature profiles for the anharmonic chain with $\omega_d=1.5$, $F_d = 10$. The temperature is flat and displays no gradient across the bulk of the system. Inset: the current is independent of system size $j \sim N^0$, which demonstrates ballistic transport in the resonant nonlinear wave regime.}
    \label{fig:temp2}
\end{figure}

In Fig.~\ref{fig:temp2}, we show the temperature and current profile for driving parameters ($\omega_d = 1.5, F_d = 10$) for which the NESS belongs to the resonant nonlinear wave regime. For system sizes below some critical value $N < N_c$ (which depends on $\omega_d$ and $F_d$), we observe a flat temperature profile in the bulk as well as a current that is independent of system size, which shows that transport in this regime is ballistic. These features are unexpected since they are characteristic of the harmonic chain and, more generally, of \textit{integrable} systems, such as the Toda lattice, connected to thermal reservoirs~\cite{zotos2001,shastry2010,chen2014,prelov2021}. That such a regime appears at any $N$ is surprising since the driving and the dissipation explicitly break translation symmetry, yet we observe a NESS with emergent translation invariance in the bulk, as evident from the temperature profile. However, upon fixing the drive parameters and varying the system size, this behavior is modified for longer chains ($N > N_c$), beyond which the system becomes chaotic and develops a temperature gradient (see Sec.~\ref{sec:mixed} for details on this regime). As such, the system resembles an unusual diffusive system with a very long (extensive) mean-free path. 

\begin{figure}[t]
     \centering
     \begin{subfigure}[b]{0.23\textwidth}
         \centering
         \includegraphics[width=\textwidth]{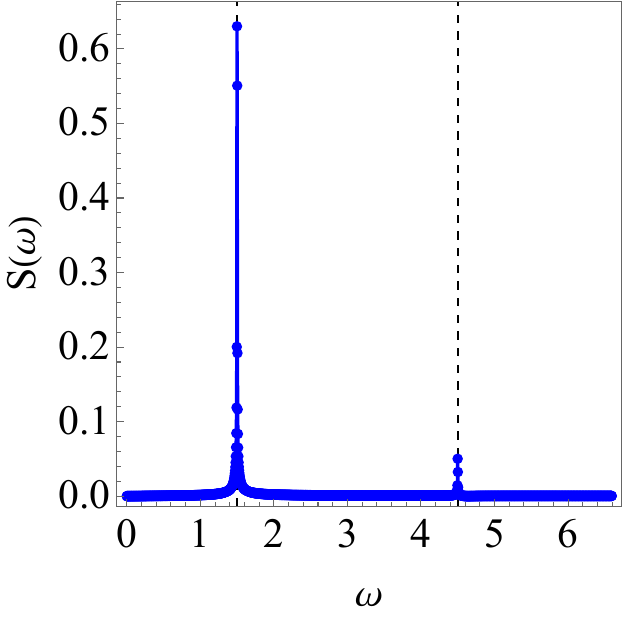}
         \caption{}
         \label{fig:anharmonicSW2}
     \end{subfigure}
     \hfill
     \begin{subfigure}[b]{0.23\textwidth}
         \centering
         \includegraphics[width=\textwidth]{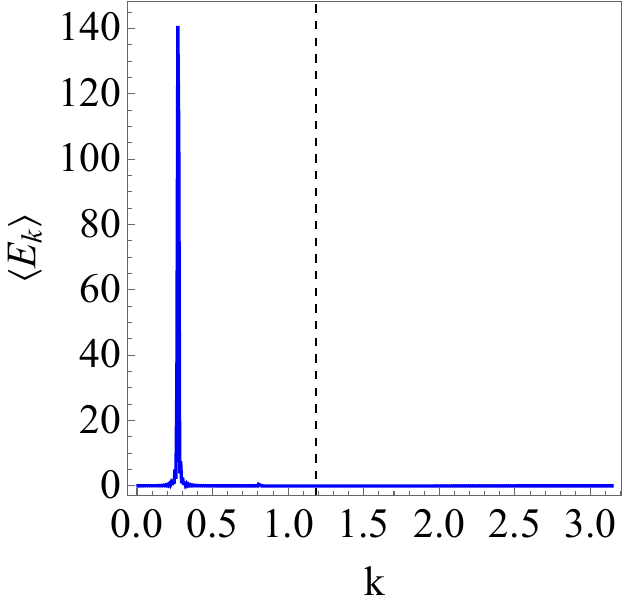}
         \caption{}
         \label{fig:anharmonicEK2}
     \end{subfigure}
     \\
     \begin{subfigure}[b]{0.23\textwidth}
         \centering
         \includegraphics[width=\textwidth]{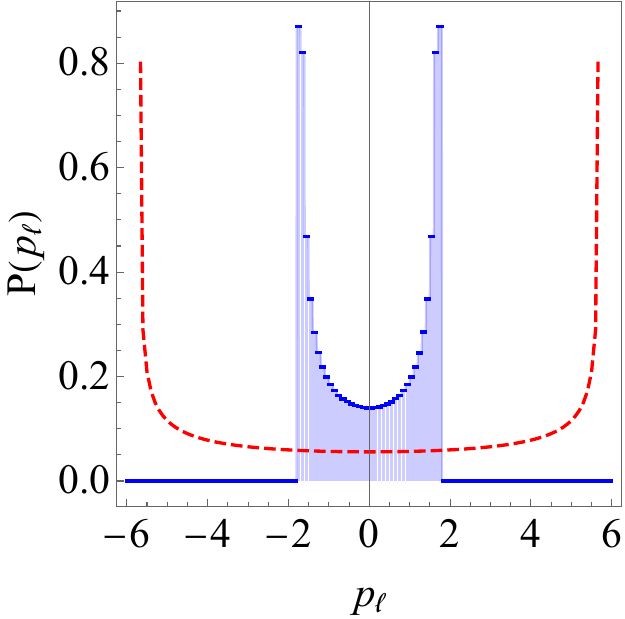}
         \caption{}
         \label{fig:anharmonicdist2}
     \end{subfigure}
     \hfill
     \begin{subfigure}[b]{0.23\textwidth}
         \centering
         \includegraphics[width=\textwidth]{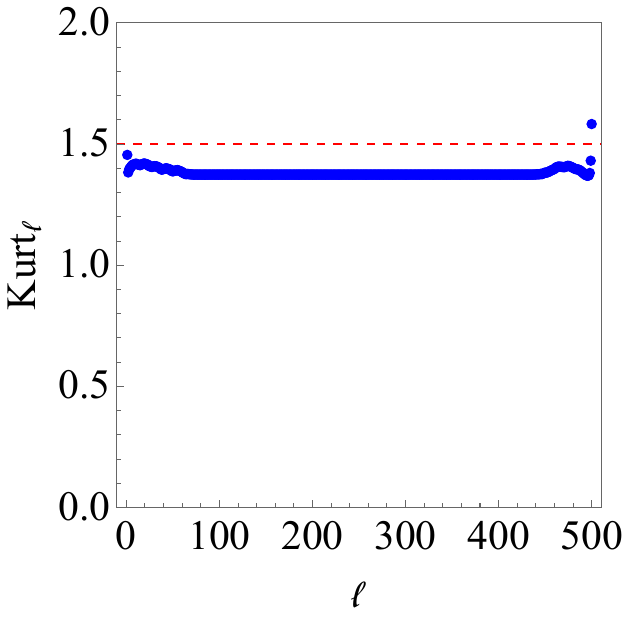}
         \caption{}
         \label{fig:anharmonickurt2}
     \end{subfigure}     
        \caption{Numerically obtained (a) power spectrum $S(\omega)$ and (b) time-averaged mode spectrum $\braket{E_k}$ in the steady-state of the anharmonic chain with $F_d = 10$, $\omega_d = 1.5$, and $N = 500$. The steady-state response shows a dominant peak at the driving frequency as well as a secondary peak at $3 \omega_d$ (dashed black lines in (a)). Only the normal mode with quasimomentum $k \approx 0$ survives in the steady-state. The dashed black line in (b) indicates the quasimomentum $k^*$ with $\omega_{k^*} \approx \omega_d$. The (c) bimodal local canonical momentum distribution (shown here for site $\ell = 245$) and (d) kurtosis imply a non-ergodic, non-thermal NESS but are distinct from those of the harmonic chain (dashed red lines).}
        \label{fig:spec2}
\end{figure}

The resonant nonlinear wave can be further characterized by its power spectrum $S(\omega)$ and time-averaged mode spectrum $\braket{E_k}$ in the non-equilibrium stationary state, which are shown in Fig.~\ref{fig:spec2}. The power spectrum in this regime shows a clear peak at the driving frequency but, in contrast with the harmonic chain, also displays a secondary peak at $\omega = 3 \omega_d$. This sub-dominant superharmonic response is a consequence of the cubic nonlinearity and, given that the linear system cannot generate higher-order harmonics, provides a crisp signature distinguishing the anharmonic chain in this regime from a harmonic chain. The approximate translation invariance of the NESS in this regime is also evident in the time-averaged spectrum of mode energies $\braket{E_k}$, which is sharply peaked close to $k = 0$ rather than at the quasimomentum $k^*$ corresponding to $\omega_{k^*} \approx \omega_d$, which would be resonant in the harmonic chain. The bimodal local canonical momentum distributions and kurtosis show that the NESS is non-thermal, but also differ from the analytic solutions obtained for the harmonic chain, as depicted in Fig.~\ref{fig:spec2}. Together, these diagnostics reveal that the resonant nonlinear wave displays ballistic transport with a \textit{non-ergodic, non-thermal current-carrying} NESS that is distinct from the steady-state for the linear system.

\begin{figure}[t]
\centering
\begin{subfigure}[b]{0.4\textwidth}
         \centering
         \includegraphics[width=\textwidth]{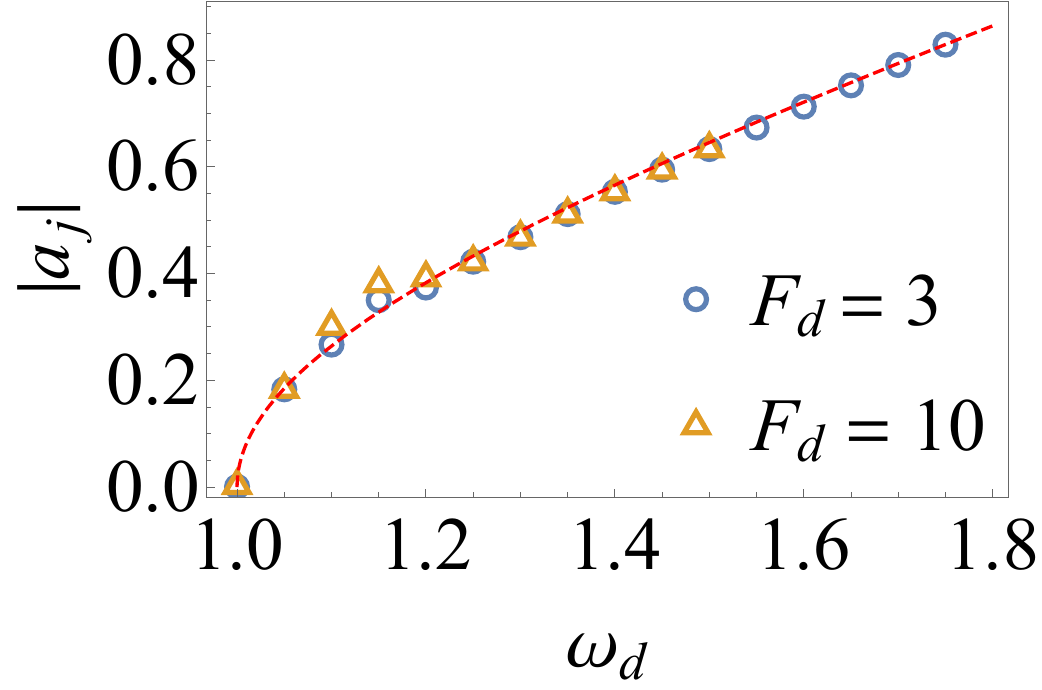}
         \caption{}
         \label{fig:matching}
     \end{subfigure}
     \\
     \begin{subfigure}[b]{0.4\textwidth}
         \centering
         \includegraphics[width=\textwidth]{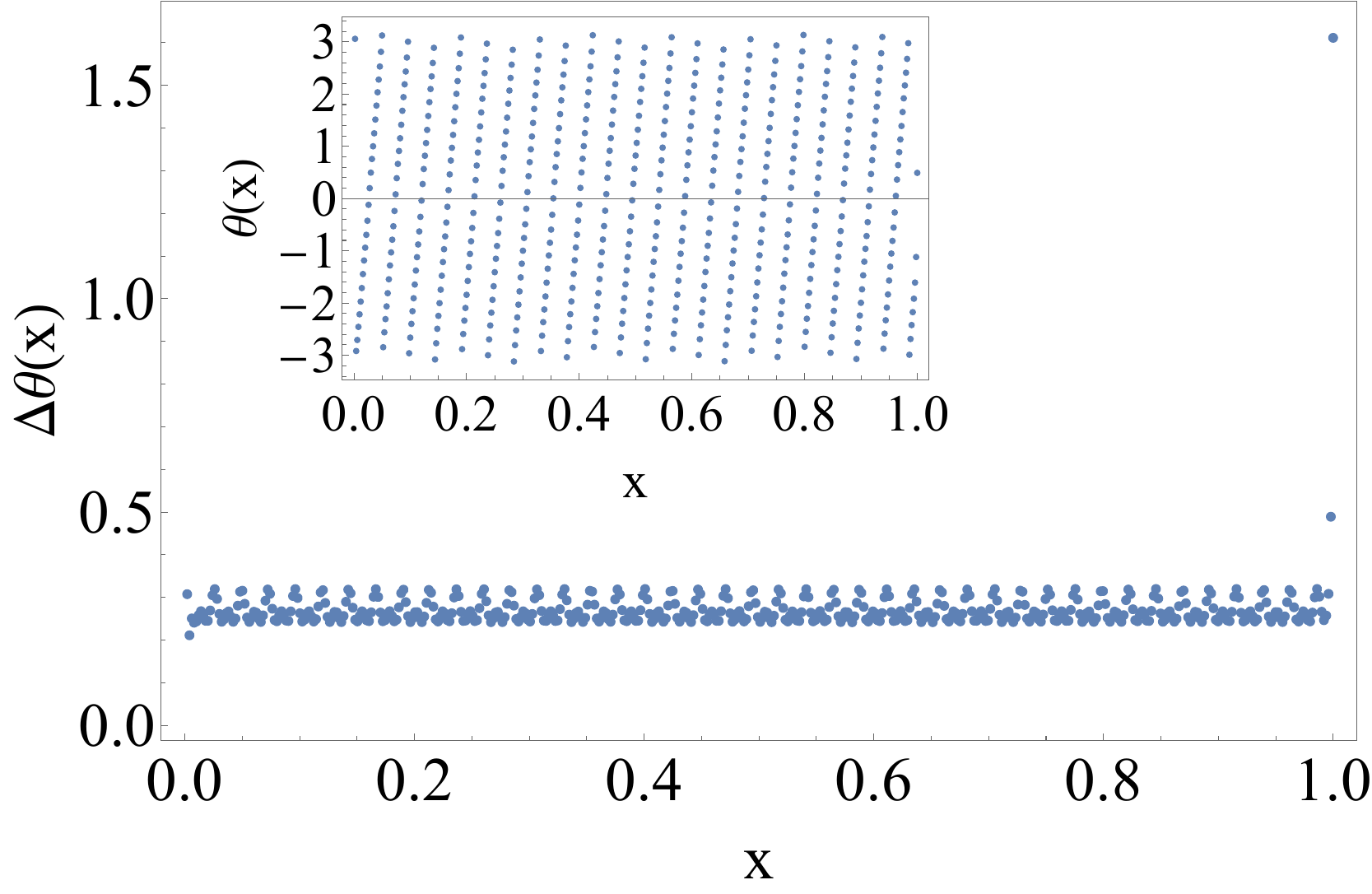}
         \caption{}
         \label{fig:phasediff}
     \end{subfigure}
    \caption{(a) The harmonic-balance solution Eq.~\eqref{eq:DBsol} (red dashed line) provides a good approximation to the numerically obtained amplitude $a_j$ (markers) in the bulk of the chain for parameter values that correspond to the resonant nonlinear wave regime. Here $j=280, \omega_d = 1.5$, and $N = 500$. (b) The phase difference $\Delta \theta(x)$ is approximately constant across the bulk of the chain. Inset: The phase $\theta(x)$ varies linearly in the bulk. Here, $\omega_d = 1.5, F_d = 10, N = 500$.}
     \end{figure}

Based on these numerical observations, we expect that the dominant behavior of the steady-state in this regime is well-approximated by a solution that is synchronized with the drive frequency (hence ``resonant'') and has most of its spectral weight localized around quasimomentum $k \approx 0$. To model the observed features of the resonant nonlinear wave, we solve for resonant solutions to the bulk Klein-Gordon dynamics (i.e. an infinite chain), which oscillate at the driving frequency, $q_j(t) = a_je^{i\omega_d t} + a_j^*e^{-i\omega_d t}$. The bulk harmonic-balance condition is given by
\begin{equation}
(-\omega_d^2 + 1 + 3|a_j|^2) a_j + \epsilon(2a_j-a_{j+1}-a_{j-1}) = 0.
\end{equation}
Provided the driving frequency $\omega_d>1$, this equation has a translation invariant solution given by
\begin{equation}
\label{eq:DBsol}
a_j \equiv \sqrt{\frac{\omega_d^2-1}{3}}.
\end{equation}
As shown in Fig.~\ref{fig:matching}, this solution is a good approximation to the numerically obtained steady-state solution for parameter values which correspond to this dynamical regime.

We note that this solution cannot be exact since it implies a vanishing bulk current and a kurtosis of 1.5, which contradicts numerics. From our numerical results, we expect that leading-order corrections to this translation invariant solution (with spectral weight concentrated entirely at $k = 0$) correspond to higher-harmonic terms which shift the spectral weight away from $k = 0$ to a small but finite value $k^*$ and are also responsible for the observed deviation of the kurtosis. In fact, we can improve our ansatz by allowing for a non-zero $k$ as follows~\cite{dharpc}: $a_j \to \alpha e^{i j k^*}$, where $k^* \equiv k^*(\omega_d)$ depends on the driving frequency. Using this ansatz, we find
\beq
\label{eq:DBsol2}
a_j = e^{ij k^*} \sqrt{\frac{\omega_d^2 - 1 - 2 \left(1 - \cos(k) \right)}{3}} \, ,
\eeq
which describes a non-chaotic nonlinear wave solution with the same amplitude on each site but with a phase that varies linearly across the chain, resulting in a constant $U(1)$ phase difference between neighboring sites. The steady-state current associated with this solution can also be calculated and is given by $j = 2 \alpha^2 \omega_d \sin(k)$, which closely matches the numerically obtained steady-state current. We also see numerically that the argument $\mathrm{arg}(a_j)$ in the steady state exhibits a near-constant shift modulo $2 \pi$ as $j \mapsto j+1$, which provides strong further evidence in favour of Eq. \eqref{eq:DBsol2} (see Fig.~\ref{fig:phasediff}). This emergent near-conservation of phase differences is strongly reminiscent of the physics of the discrete nonlinear Schr{\"o}dinger equation at low temperature~\cite{Kulkarni_2015,Mendl_2015}.

We dub this solution a ``resonant nonlinear wave'' because it is analogous to the nonlinear wave solutions that are known to arise in other chaotic classical lattice models in the absence of driving. However, for such models, which include the discrete nonlinear Schr{\"o}dinger equation~\cite{kivshar1993peierls} and the classical XXZ chain~\cite{Lakshmanan_2011}, 
the nonlinear wave is an exact solution. Although for the Klein-Gordon chain, our nonlinear wave solution Eq.~\eqref{eq:DBsol2} is only an approximate solution to the bulk dynamics (because it invokes the harmonic balance approximation), the robustness of our numerical observations suggests that Eq.~\eqref{eq:DBsol2} may be close to an exact periodic orbit of the bulk Klein-Gordon system. It for this reason that we refer to the solution Eq.~\eqref{eq:DBsol2} as a ``resonant nonlinear wave''. To the best of our knowledge, this phenomenon has not been reported before in anharmonic chains, though it is akin to the
nonlinear normal modes that can arise in Fermi-Pasta-Ulam-Tsingou (FPUT) chains without driving and dissipation~\cite{chechin2012,peng2022}.


We now consider linear (modulational) stability of the solution Eq.~\eqref{eq:DBsol2}, by considering time-dependent perturbations of the form $q_j(t) = (a_j+b_j(t))e^{i\omega_d t} + (a_j^*+b_j^*(t))e^{-i\omega_d t}$, with $|b_j| \ll a_j$ and $|\dot{b}_j| \ll \omega_d |b_j|$. Within the rotating wave approximation, this yields
\begin{equation}
2i\omega_d \dot{b}_j = (1-\omega_d^2)(b_j+b_j^*)-\epsilon(2b_j-b_{j+1}-b_{j-1}).
\end{equation}
Solving for normal modes $b_j = u_j e^{\lambda t}+ v_j^* e^{\lambda^*t}$ of this equation and Fourier transforming in space, we find that the exponents $\lambda(Q)$ at wavenumber $Q$ satisfy
\begin{equation}
\lambda(Q)^2 = - \frac{2\epsilon \sin^2{Q/2}}{\omega_d^2}\left(2\epsilon \sin^2{Q/2} + \omega_d^2 - 1\right) < 0
\end{equation}
for $\omega_d>1$, implying that the resonant nonlinear wave solution is stable to weak modulations. The above analytical prediction of linear stability is consistent with the numerical observation of vanishing Lyapunov exponents in this regime (see Sec.~\ref{sec:lyapunov}).

Despite its linear stability, we find numerically that, at fixed system size, the resonant nonlinear wave undergoes a nonlinear instability as the driving force is increased past a certain threshold $F_{c2}$, as shown in Fig.~\ref{fig:JvsFmain}. Past this threshold (which depends on the drive frequency $\omega_d)$, the single-mode single-frequency approximation breaks down due to the broadening of the $k \approx 0$ peak and the system enters a chaotic regime, which we discuss in the following section. We also observe numerically that for fixed driving parameters, the resonant nonlinear wave decays in real space in sufficiently long chains, at a wavelength $Q^{-1} \propto N$ (see Fig.~\ref{fig:Scaling2}). We attribute this breakdown to a bulk nonlinear instability that is not captured by linear response. While our results indicate that the resonant nonlinear wave will disappear in the thermodynamic limit, it should nevertheless appear as a distinct dynamical regime at system sizes accessible to current circuit QED experiments operating near the semiclassical limit.


\subsection{Spatial Ballistic-to-Diffusive Crossover with Diffusive Transport}
\label{sec:mixed}

The final dynamical regime that we find for the boundary-driven dissipative discrete anharmonic chain is obtained once the driving amplitude is increased past the second threshold $F_d > F_{c2}$ with $N$ fixed or when the chain length is increased past $N_c$ for a fixed drive. As discussed in the previous section, the resonant nonlinear wave is linearly stable and so the onset of this instability is most likely due to a bulk nonlinear instability, which should be visible as a broadening of the $k \approx 0$ peak observed in the time-averaged mode spectrum $\braket{E_k}$. While we do not have an analytic understanding of this onset, it is likely that the nonlinear interactions between the resonant nonlinear wave solution Eq.~\eqref{eq:DBsol} and the phonon modes of the underlying harmonic lattice are responsible for this broadening, which causes the resonant nonlinear wave to decay~\cite{johansson2000}. As we now discuss, this strongly driven regime corresponds to a highly unusual NESS that is only locally thermalized over part of the chain and is not locally thermal elsewhere, seeming to resemble a ballistic-to-diffusive crossover occurring in \textit{space}, rather than time. 

\begin{figure}[t]
    \centering
    \includegraphics[width=0.5\textwidth]{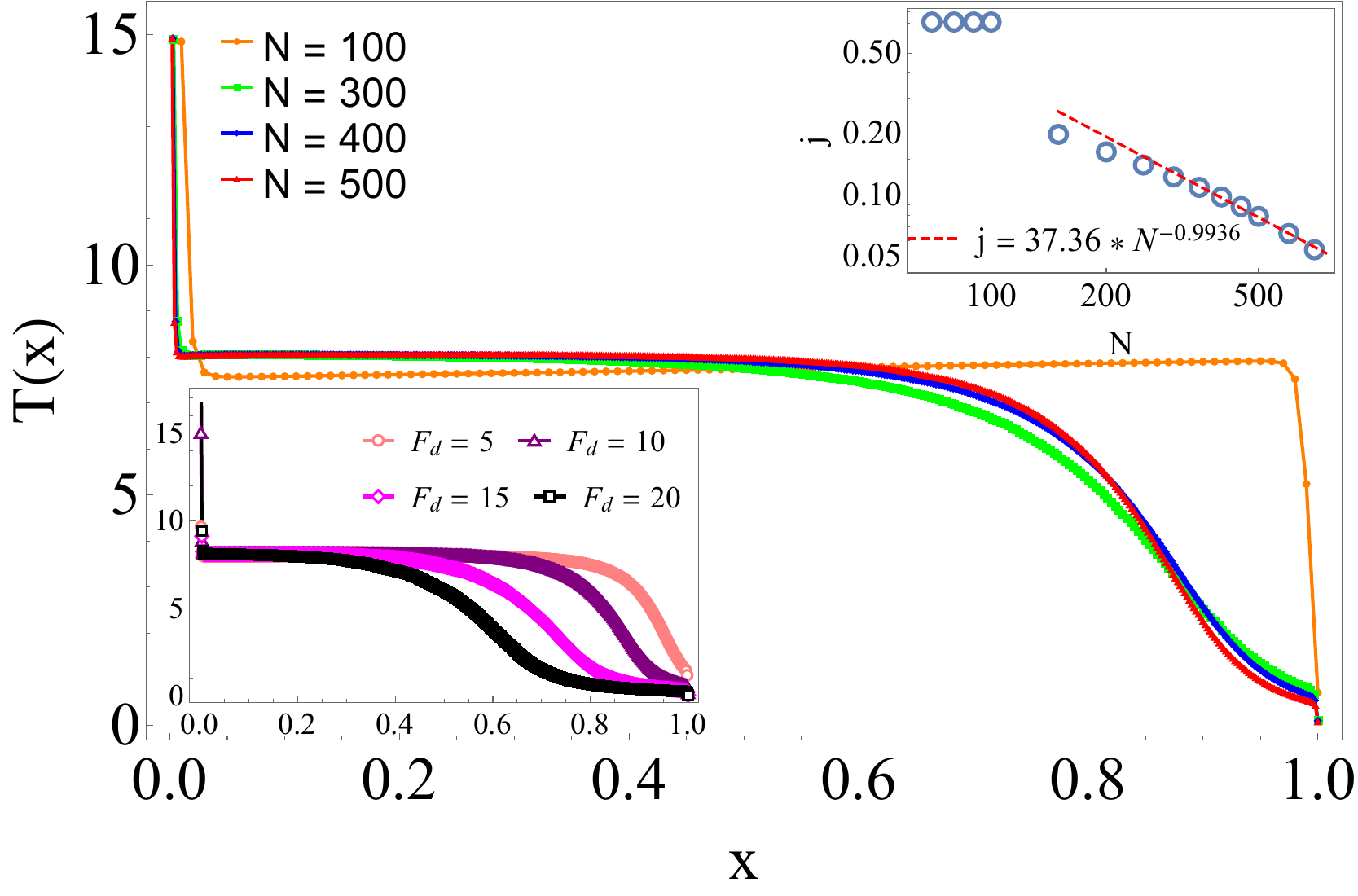}
    \caption{Temperature profiles for the anharmonic chain with $\omega_d = 2.0$, $F_d = 10$. The steady-state in this regime is spatially inhomogeneous, with a temperature profile that is initially flat but then develops a gradient further along the chain. Upper inset: the steady-state energy current decays with system size as $\sim N^{-1}$, indicating that transport is diffusive in this regime. Markers denote numerical data points and the dashed red line indicates the corresponding fit. Lower inset: the spatial extent over which the system displays a temperature gradient increases with the drive amplitude $F_d$, which shows that this is not merely a boundary effect (here $N = 500$).}
    \label{fig:temp3}
\end{figure}

The NESS averaged temperature profile for driving parameters corresponding to this regime ($\omega_d = 2.0, F_d = 10.0$) is shown in Fig.~\ref{fig:temp3} for various system sizes. For system sizes above $N_c$ (below which we observe a resonant nonlinear wave), the temperature profile is highly nonlinear, with a temperature gradient that vanishes for an extensive number of sites on the left part of the chain (closer to the driven boundary), but is non-trivial for an extensive number of sites on the right part of the chain (closer to the undriven boundary). Na{\"i}vely, one would expect a profile similar to the thermal regime (see Sec.~\ref{sec:thermal}), where the entire bulk of the chain supports a thermal gradient, but the interplay between inhomogeneous driving, dissipation, and nonlinearity here results in an atypical steady state in which energy is predominantly localized in the initial section of the chain. 

We emphasize that the presence of a non-trivial gradient along the latter part of the chain cannot be regarded as a boundary or ``skin" effect. Firstly, as evident from Fig.~\ref{fig:temp3}, the spatial extent of this region increases with system size $N$ and is thus extensive, in contrast with a boundary layer which should possess a depth that remains finite as $N \to \infty$. Secondly, the lower inset of Fig.~\ref{fig:temp3} illustrates that the width of this region grows with the driving strength, which further suggests that the observed features are not merely a boundary effect. While we did not obtain a convergent NESS for larger driving amplitudes than those displayed for $N = 500$, based on these results it is reasonable to expect that as $F_d$ is increased even further, the entire chain will eventually develop a temperature gradient\footnote{We note however that in the limit $F_d \to \infty$, the first driven site will become dynamically decoupled from the rest of the chain, leading to a vanishing energy current and trivial dynamics for all other sites. However, we do not discuss this regime here as our numerics did not converge to a stationary state for $N = 500$ beyond $F_d \approx 25$.}. Finally, we observe normal \textit{diffusive} transport in this regime (see upper inset of Fig.~\ref{fig:temp3} and also Fig.~\ref{fig:Scaling2}), which is consistent with this being a bulk phenomenon; were it a boundary effect, we would have observed ballistic transport associated with the flat temperature profile displayed by the resonant nonlinear wave. 

\begin{figure}[t]
     \centering
     \begin{subfigure}[b]{0.23\textwidth}
         \centering
         \includegraphics[width=\textwidth]{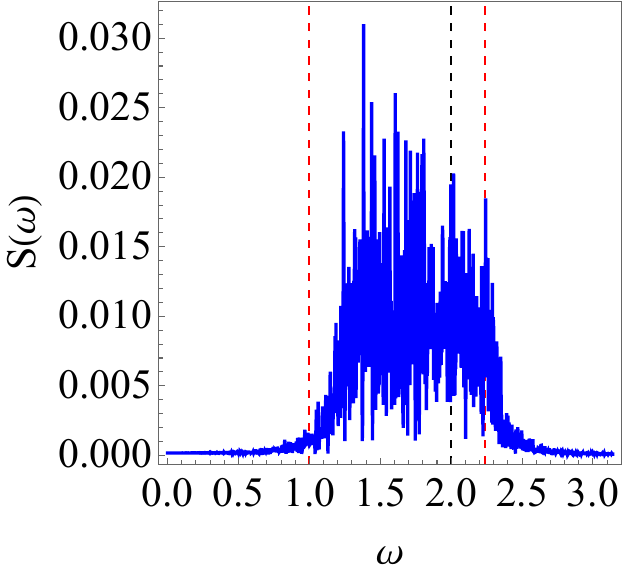}
         \caption{}
         \label{fig:anharmonicSW3}
     \end{subfigure}
     \hfill
     \begin{subfigure}[b]{0.23\textwidth}
         \centering
         \includegraphics[width=\textwidth]{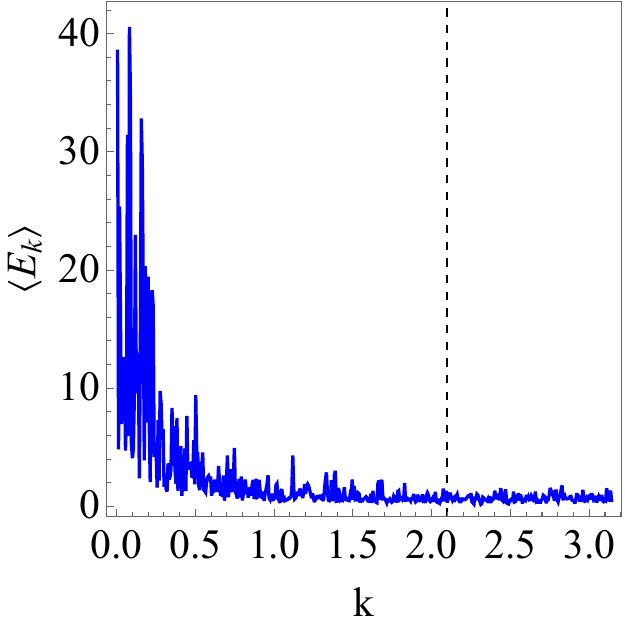}
         \caption{}
         \label{fig:anharmonicEK3}
     \end{subfigure}
         \\
     \begin{subfigure}[b]{0.23\textwidth}
         \centering
         \includegraphics[width=\textwidth]{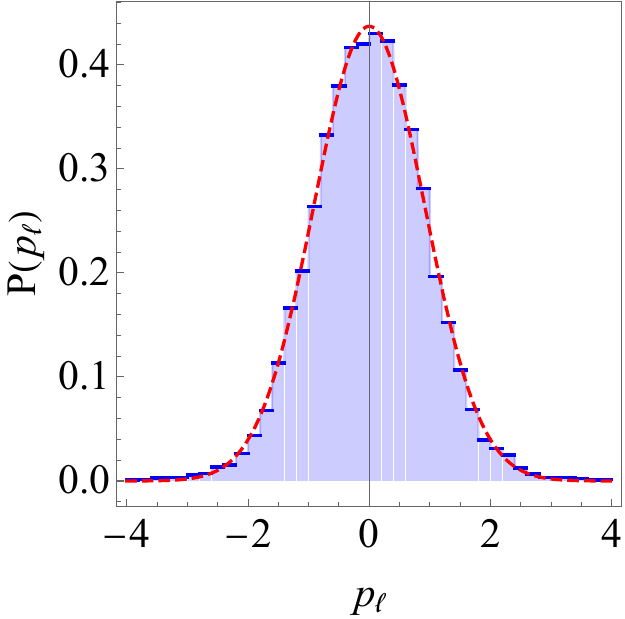}
         \caption{}
         \label{fig:anharmonicdist3}
     \end{subfigure}
     \hfill
     \begin{subfigure}[b]{0.23\textwidth}
         \centering
         \includegraphics[width=\textwidth]{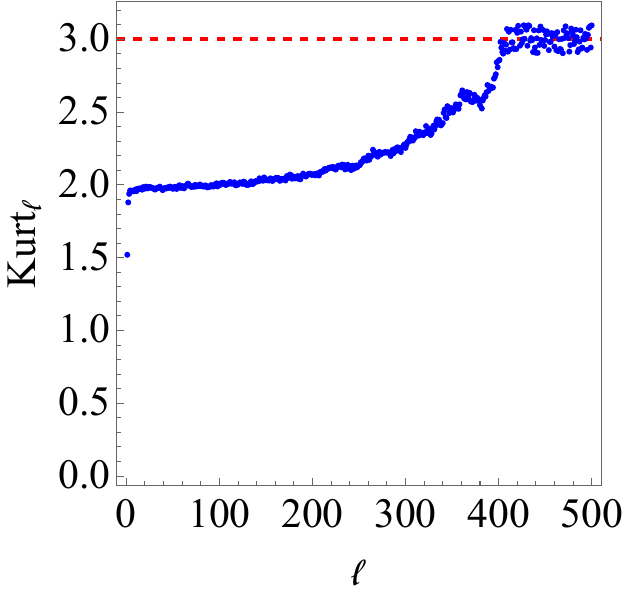}
         \caption{}
         \label{fig:anharmonickurt3}
     \end{subfigure} 
        \caption{Numerically obtained (a) power spectrum $S(\omega)$ and (b) time-averaged mode spectrum $\braket{E_k}$ in the steady-state of the boundary-driven dissipative anharmonic chain with $F_d = 10$, $\omega_d = 2.0$, and $N = 500$. The emission spectrum is broadband and extends outside the upper harmonic band edge, while the mode spectrum shows a broadening around the translation invariant quasimomentum $k \approx 0$. Dashed red lines in (a) indicate the harmonic-band edges, while dashed black lines in (a), (b) indicate the driving frequency $\omega_d$ and the quasimomentum corresponding to $\omega_{k^*} \approx \omega_d$. (c) The local canonical momentum distribution (shown for $\ell = 480$) matches the Maxwell-Boltzmann distribution Eq.~\eqref{eq:MBdist} (in red) with temperature set by the local temperature $T_\ell$ for sites $\ell \gtrsim 415$. (d) The kurtosis matches the thermal expectation value (red dashed line) only for sites $\ell \gtrsim 415$.}
        \label{fig:spec3}
\end{figure}

We can characterize this dynamical regime further through the probes defined in Sec.~\ref{sec:probes}, namely the power spectrum $S(\omega)$ and the time-averaged mode spectrum $\braket{E(k)}$ in the steady-state (see Fig.~\ref{fig:spec3}). While the emission spectrum is broadband as in the thermal regime, here we also observe a non-trivial response for frequencies above the upper harmonic band edge, which demonstrates that this spectrum is a consequence of nonlinear effects beyond multiply-excited normal modes. As expected, the bulk nonlinear instability of the resonant nonlinear wave solution appears as a broadening of the $k \approx 0$ peak in the mode spectrum. As $F_d$ is increased, we find that the growth in the spatial extent of the chain with a temperature gradient is reflected in the excitation of higher $k$-modes, indicating that full ergodicity in the space of normal modes is restored once the entire chain develops a temperature gradient. 

Figs.~\ref{fig:anharmonicdist3} and~\ref{fig:anharmonickurt3} also show the canonical local momentum distribution (for site $\ell = 480$ in an $N = 500$ site chain) and the steady-state averaged kurtosis across the chain respectively. The latter diagnostic reveals that only sites located within the spatial region with a non-trivial temperature gradient are in thermal equilibrium. Closer inspection of the local momentum distributions reveals a real-space crossover from the bimodal behavior observed in the resonant nonlinear wave to a Maxwell-Boltzmann distribution consistent with local thermalization. In other words, the chain displays \textit{partial thermalization in real space}, by which we mean a crossover in real-space from a region that is not in local thermal equilibrium to a region that is in local thermal equilibrium. We note that these features are markedly different from those of the discrete Klein-Gordon chain in contact with thermal reservoirs, which supports diffusive transport but a NESS that is fully thermalized in the bulk.


\subsection{Non-local Lyapunov exponent}
\label{sec:lyapunov}

\begin{figure*}[t]
     \centering
     \begin{subfigure}[b]{0.46\textwidth}
        \hspace{-0.5in}
         \centering
         \includegraphics[width=\textwidth]{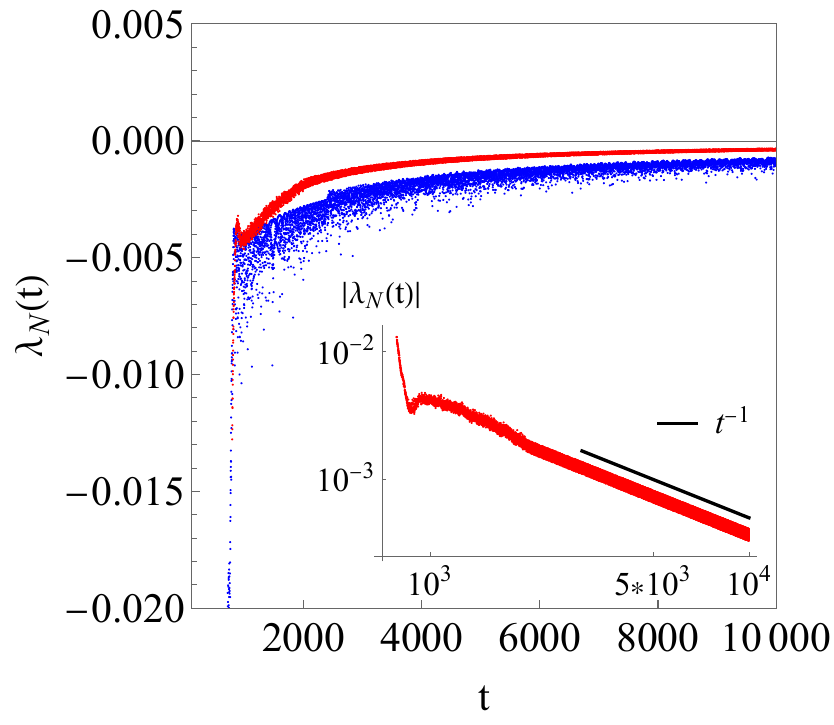}
         \caption{}
         \label{fig:lyap1}
     \end{subfigure}
     \,
    \begin{subfigure}[b]{0.43\textwidth}
    \centering
    \includegraphics[width=\textwidth]{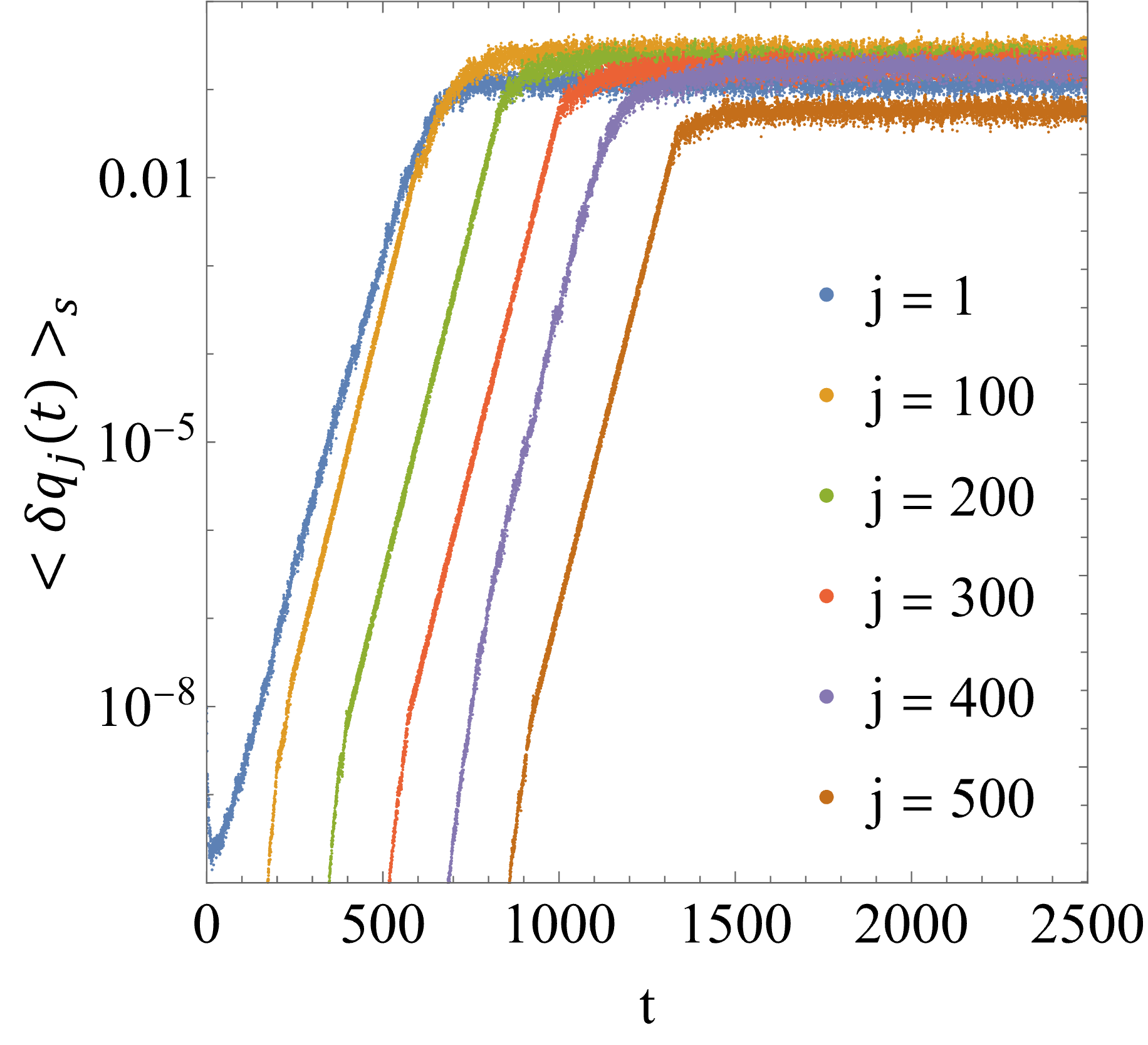}
    \caption{}
    \label{fig:lyap2}
    \end{subfigure}
    \hfill \\
     \begin{subfigure}[b]{0.5\textwidth}
         \centering
         \includegraphics[width=\textwidth]{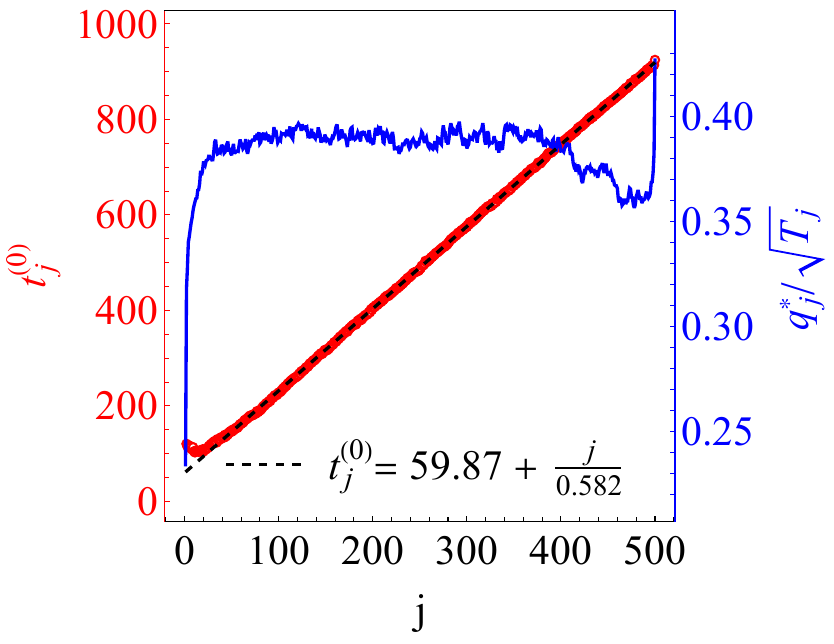}
         \caption{}
         \label{fig:lyap3}
     \end{subfigure}
          \,\,\,
     \begin{subfigure}[b]{0.42\textwidth}
         \centering
         \includegraphics[width=\textwidth]{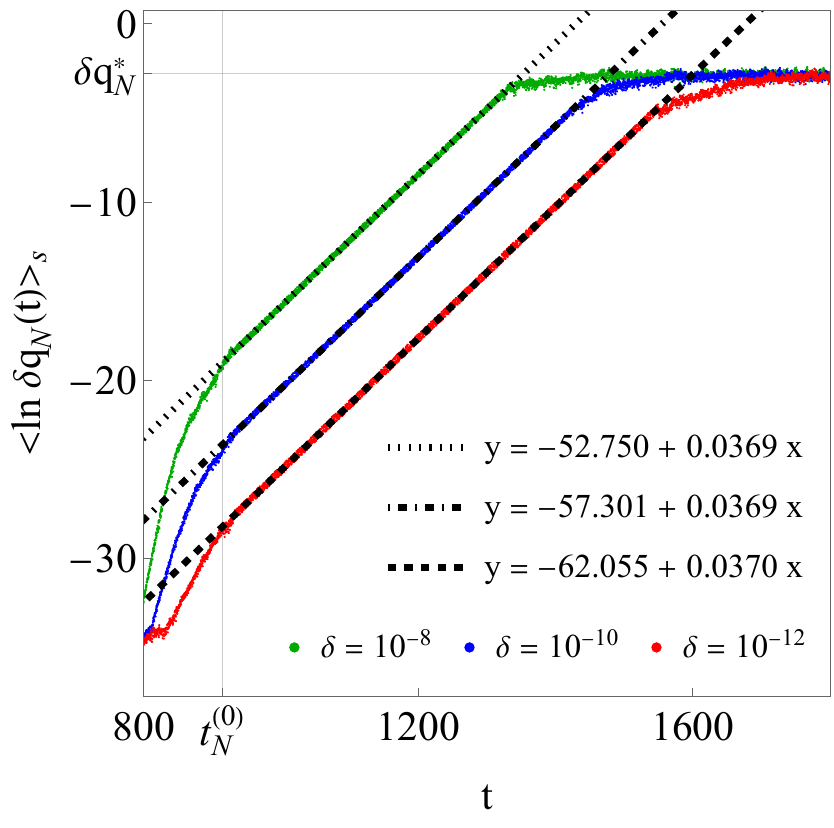}
         \caption{}
         \label{fig:lyap4}
     \end{subfigure}
        \caption{(a) The non-local Lyapunov exponent $\lambda_N(t)$ Eq.~\eqref{eq:ftle} is negative $\lambda(t) < 0 \, \forall \, t$ for the harmonic chain (in blue; $F_d = 10, \omega_d = 2.0$) and for the anharmonic chain in the resonant nonlinear wave regime (in red; $F_d = 10, \omega_d = 1.5$). In each case, $\lambda(t) \sim 1/t$ at late-times as shown on a log-log scale in the inset (the harmonic case is omitted for clarity). In the thermal regime of the anharmonic chain ($F_d = 1, \omega_d = 1.8$), (b) the deviation $<\delta q_j(t)>_s$ (on a log scale) at different sites displays a region of exponential growth at intermediate times. (c) The numerically obtained time $t_j^{(0)}$ at which the perturbation reaches site $j$ (red dots) agrees with the ballistic spread expected for a chaotic system. The dashed black line is the best linear fit. The saturation value $q_j^*$ at each site is set by the local temperature $T_j$, as indicated by the blue data points. (d) The steady-state averaged deviation $<\ln(\delta q_N(t))>_s$ on the last site grows exponentially from time $t_N^{(0)}$ until it saturates to its maximum value $\delta q_N^*$. The numerical results show excellent agreement with the behavior expected for a bounded chaotic system (see Eq.~\eqref{eq:lyap}). The best fits (black lines) yield a non-local Lyapunov exponent $\lambda_N \approx 0.037$, independent of the initial perturbation $\delta$. For all cases shown, $N = 500$ and averaging is performed over 100 initial conditions drawn from the respective steady-state distributions. The perturbation strength is $\delta = 10^{-8}$ for (a), (b), and (c).}
        \label{fig:lyapunov}
\end{figure*}

We now turn to the non-local Lyapunov exponent introduced in Sec.~\ref{sec:probes}, which provides a measure of chaos at short and intermediate time-scales, and study this quantity for the different non-equilibrium steady states of the boundary-driven dissipative Klein-Gordon chain. We note first that the NESS in all of the distinct regimes discussed above is bounded in space and in momentum; this is most clearly seen from the local canonical momentum distributions. Thus for any small perturbation strength $\delta$ (equivalently, $\phi$) the deviation $\delta q_j(t)$ between the perturbed and unperturbed trajectories will eventually saturate, such that $\lambda_j(t) \sim 1/t$ at late-times. Nonetheless, we find that the intermediate time behavior of $\lambda_N(t)$ provides a sharp measure of chaos (or its absence) in the NESS. 

Fig.~\ref{fig:lyapunov} shows the non-local Lyapunov exponent for the different dynamical regimes discussed in preceding sections. In each case, we prepare two copies of the system in identical initial states picked randomly from the steady-state distribution and add an infinitesimal phase shift $\phi$ to the external drive in one copy. As discussed in Sec.~\ref{sec:probes}, after a short time this results in a deviation $\delta$ between the positions $q_1^{(A)}$ and $q_1^{(B)}$ of the first oscillator in the two copies; we set the time at which this occurs to be $t = 0$. The two copies are then allowed to evolve independently and an average over the steady-state distribution is taken in order to calculate $\lambda_N$ from Eqs.~\eqref{eq:ftle}-\eqref{eq:lyap}.

For the harmonic chain and in the quasilinear regime of the anharmonic chain (A), which both display regular, non-chaotic dynamics, $\lambda_N(t) < 0\, \forall \, t$ and $\lambda_N(t) \sim 1/t$ for large $t$, as shown in Fig.~\ref{fig:lyap1}. Similarly, $\lambda_N(t)$ in the resonant nonlinear wave regime (C) is also negative at all times and approaches zero from below as $1/t$. In either case, we do not observe any dependence on the perturbation strength $\delta$. The lack of any region of exponential growth supports our earlier conclusion that the dynamics in these regimes is regular and non-chaotic. Thus $\lambda_N(t)$ does not qualitatively differentiate between these dynamically distinct regimes, but does reveal their shared non-ergodicity and similarity with integrable systems~\cite{dhar2018}.

The results for the thermal regime (B), which displays ergodicity in the space of normal modes, are shown in Figs.~\ref{fig:lyap2}-\ref{fig:lyap4}. After an early-time decrease, $\delta q_1(t)$ shows power-law growth until a time $t_1^{(0)}$, after which it grows exponentially starting from a value $\approx \delta$ and eventually saturates to $q_1^*$. We similarly observe in Fig.~\ref{fig:lyap2} that $\delta q_j(t)$ displays a clear region of exponential growth between time $t_j{(0)}$ and $t_j^*$. The former is the time it takes the initial perturbation to reach the $j^{th}$ site i.e., $\delta q_j(t_j^{(0)}) = \delta$ while the latter is the ($\delta$ dependent) saturation time. We note that the saturation value $q_j^*$ is site-dependent and, as we discuss shortly, is set by the local temperature $T_j$. 

Consistent with chaotic behavior, we observe ballistic propagation of the perturbation, which reaches site $j$ at a time $t_j^{(0)} = t_1^{(0)} + j/v_b$. Fig.~\ref{fig:lyap3} shows a plot of $t_j^{(0)}$ versus $j$, with the dashed black line indicating the best linear fit. From the inverse slope, we find that the butterfly velocity, or the speed of chaos propagation, is given by $v_b = 0.582$. We have verified that this value is independent of the strength of the initial perturbation $\delta$ (equivalently $\phi$). Fig.~\ref{fig:lyap3} also shows that the saturation value $q_j^*$, normalized by the square root of the local temperature $T_j$, is approximately uniform across the bulk of the chain. This confirms that ``boundedness'' of the dynamics in this driven system is a consequence of its tending to a state of local thermal equilibrium. The latter property implies that at sufficiently long times, the characteristic fluctuations of $q_j$ and $p_j$ at a given site are set by the scale $\sim \sqrt{T}_j$ of thermal fluctuations at that site.

Finally, in Fig.~\ref{fig:lyap4} we plot the  deviation on the last site (averaged over the steady-state distribution) for different values of the initial perturbation $\delta$. Once the initial perturbation on this site grows to $\delta$ at time $t_N^{(0)}$, $\delta q_N(t)$ shows a clear region of exponential growth at intermediate times. As expected, the saturation time grows with decreasing $\delta$ and, consequently, so does the time scale over which exponential increase is visible. We find remarkably good agreement between the numerical results and the expected behavior Eq.~\eqref{eq:lyap} in this regime, from which we can extract the non-local Lyapunov exponent $\lambda_N$ for the NESS. From the best linear fits, shown in Fig.~\eqref{fig:lyap4}, we find $\lambda_N = 0.037$ with no significant dependence on $\delta$. We likewise find a positive non-local Lyapunov exponent in the partially thermalized regime (D) (not shown here). 

Thus, the non-local Lyapunov exponent $\lambda_N(t)$ provides a clear and unambiguous non-local measure of chaos in the NESS: for regular, non-chaotic dynamics, we find that $\lambda_N(t) \leq 0 \, \forall \, t$ while chaotic systems display exponential growth at intermediate times from which we can obtain a positive exponent $\lambda_N$. We emphasize that this quantity is not intended to distinguish between different non-ergodic regimes or to differentiate the thermal regime with anomalous transport from the partially thermalized regime with diffusive transport; rather, its purpose is to diagnose chaos in the bulk of the system through a measurement that only probes its boundaries.


\section{Conclusions}
\label{sec:cncls}

We have investigated the classical dynamics of boundary-driven, dissipative, discrete Klein-Gordon chains and revealed that they support multiple physically distinct non-equilibrium stationary states. In contrast to the oft-studied scenario of anharmonic chains driven stochastically at their boundaries by thermal reservoirs, for which the NESS is typically locally thermal and displays diffusive transport, we find that a coherent, periodic drive allows a much wider range of possibilities for the long-time dynamics. Specifically, we have shown that for a fixed chain length, varying the driving parameters can lead to a NESS that is either non-thermal or is in local thermal equilibrium (either along part of the chain or along the entire chain), with energy transport that can be either ballistic, diffusive, or anomalous depending on the state under consideration.

Besides these bulk probes, we have also discussed how two experimentally relevant quantities---the power spectral density $S(\omega)$ and the non-local Lyapunov exponent $\lambda_N(t)$---provide clear probes of thermalization and chaos in this system. 
Important questions remain for the specific model considered here: firstly, a more thorough analysis is required to reveal whether there are any sharp transitions in this model as a function of varying drive amplitude (at fixed system size and drive frequency) or if these are instead smooth crossovers. While our numerics suggest a transition between the resonant nonlinear wave and the partially thermalized regime, it remains unclear whether this transition is sharp in the appropriate large-system limit.

Similarly, the physics that controls the instability of the resonant nonlinear wave solution to the partially thermalized regime remains to be understood. In order to isolate the effects of the periodic drive, we set the reservoir temperatures to zero throughout this paper---it would be interesting to investigate whether there is a range of bath temperatures over which the non-thermal behavior induced by the periodic drive remains stable. The resonant nonlinear wave solution in particular is expected to be unstable to the incoherent driving induced by thermal reservoirs, but there likely exists a crossover region at low bath temperatures where this solution is approximately valid. 
 
Our work invites further study of classical systems with periodic driving and dissipation that are spatially localized at the boundaries. For instance, while integrable systems coupled at their boundaries to thermal reservoirs typically exhibit ballistic transport~\cite{dhar2018}, the effect of a coherent boundary drive on these systems is largely unexplored. Along these lines, it would be interesting to study whether approximately integrable systems such as the FPUT chain at low energies, which are known to display slow relaxation towards equipartition of energy amongst normal modes~\cite{matsuyama2015,danieli2019}, harbor non-equilibrium stationary states with anomalous dynamics in the boundary-driven dissipative setting. Given that many nonlinear classical systems can be realized within the framework of nonlinear electric transmission networks~\cite{kengne2022}, our work also provides motivation for the experimental study of boundary-driven dissipative classical arrays.

Returning to the primary motivation for this study, a natural question is whether our results can provide insight into the dynamics of interacting circuit QED arrays. Specifically, the classical system considered here recovers the phenomenology of driven-dissipative quantum chains, which display a transition from a transmitting regime at low driving amplitudes to a non-transmitting regime at higher driving strengths~\cite{fitzpatrick2017,fedorov2021}. That a similar transition is found (albeit with additional intervening regimes) in our classical model suggests that the experimentally observed transition is operating in the semiclassical regime of large photon number where quantum fluctuations are strongly suppressed. 

A more thorough analysis of the corresponding theoretical models is thus required to shed light on the role of quantum effects in the experimentally observed transition. For circuit QED experiments that operate near the semiclassical limit, our results also suggest the possibility of tuning between truly far-from-equilibrium steady states and those close to local thermal equilibrium as a function of the driving strength. While experiments on large arrays do not currently have access to individual lattice sites in the bulk, signatures of these distinct dynamical regimes can nevertheless be probed through appropriate modifications of the diagnostics discussed here. Meanwhile, understanding the dynamics of driven, dissipative quantum systems \textit{away} from the semiclassical limit continues to pose an exciting scientific challenge.


\section{Acknowledgments}

We are especially grateful to David Campbell, Abhishek Dhar, Manas Kulkarni, Joel Lebowitz, and Biao Lian for illuminating discussions and comments on the draft. A.P. and S.L.S would also like to thank Alicia Koll\'ar for stimulating discussions on circuit QED experiments and open quantum systems, which led to the initiation of this work. V.B.B. thanks W. Benalcazar and F. Schindler for helpful discussions on related topics. Computational resources were provided by the Feynman and Della clusters at Princeton University. A.P. and V.B.B. acknowledge support from the Princeton Center for Theoretical Science. This material is based upon work supported by the U.S. Department of Energy, Office of Science, Office of High Energy Physics under Award Number DE-SC0009988. This work was supported by a Leverhulme Trust International Professorship grant number LIP-202-014 (S.L.S). For the purpose of Open Access, the author has applied a CC BY public copyright licence to any Author Accepted Manuscript version arising from this submission.


\bibliography{library}


\appendix

\section{Generalized Langevin Equations for the System}
\label{sec:eliminate}

In this Appendix, we derive the generalized Langevin equations describing a many-body system of anharmonic oscillators which is driven at one end and experiences dissipation at its boundaries. Specifically, we start from the microscopic system-bath Hamiltonian discussed in Sec.~\ref{sec:models} and discuss the various approximations under which the system undergoes Markovian dynamics after the bath is eliminated.  

In what follows, we will assume that the system is uncoupled from the baths at some early time $t_0$ (with the initial conditions of the system and baths specified at $t=t_0$) and that the driving force is switched off for $t<t_0$. In particular, we assume here that all system and bath oscillators are initially at rest: $q_j(t_0) = p_j(t_0) = 0$ and $q_{j,\alpha}(t_0) = p_{j,\alpha}(t_0) = 0$. At time $t_0$, the system and reservoirs are coupled and the driving force is switched on, such that the evolution of the system plus reservoirs is governed by the total Hamiltonian Eq.~\eqref{eq:totalH}. We will also require a continuous distribution of bath modes i.e., $N_\alpha \to \infty$ which is necessary in order to avoid Poincar\'e recurrences~\cite{mazur60,weiss2012quantum} and to accurately capture dissipative effects, where energy flows irreversibly from the system into the bath modes. Since we are interested in the steady-state properties of the system, we will also take the limit $t_0 \to -\infty$. Finally, since we are interested here in understanding the effects of Markovian dissipation on the system of interest, we will take the continuum string limit for the harmonic chain, resulting in an Ohmic spectral density.

To proceed, it will be convenient to work in the normal mode basis of bath modes. We re-write the bath Hamiltonians as
\beq
H_{B,\alpha} = \frac{1}{2} \bm{p}_\alpha^T \bm{M}_\alpha^{-1} \bm{p}_\alpha + \frac{1}{2} \bm{q}_\alpha^T \bm{K}_\alpha \bm{q}_\alpha \, ,
\eeq
with $\bm{q}_\alpha = \left(q_{1,\alpha},\cdots,q_{N_\alpha,\alpha} \right)^T$, $\bm{p}_\alpha = \left(p_{1,\alpha},\cdots,p_{N_\alpha,\alpha} \right)^T$, the mass matrix $\bm{M}_\alpha = m_\alpha \bm{I}$, and $\bm{K}_\alpha$ are the tri-diagonal inter-particle coupling matrices. We can then transform to the normal mode coordinates and momenta
\beq
\bm{Q}_\alpha = \sqrt{m_\alpha} \bm{\Lambda}_\alpha \bm{q}_\alpha, \quad \bm{P}_\alpha = \frac{1}{\sqrt{m_\alpha}} \bm{\Lambda}_\alpha \bm{p}_\alpha \, ,
\eeq
where $\bm{\Lambda}_\alpha$ is an orthogonal matrix which diagonalizes the bath Hamiltonian: $\bm{\Lambda}_\alpha \bm{K}_\alpha \bm{\Lambda}_\alpha^T = m_\alpha \bm{\Omega}_\alpha^2$. The elements of the diagonal matrix $\bm{\Omega}^2_\alpha$ give the normal mode frequencies of the bath: $(\Omega^2_\alpha)_{ij} = \omega^2_{j,\alpha} \delta_{j,j} (j=1,\cdots,N_\alpha)$. Hence, the bath Hamiltonians are re-expressed as
\beq
H_{B,\alpha} = \frac{1}{2} \sum_{j=1}^{N_\alpha}\left[ P_{j,\alpha}^2 + \omega_{j,\alpha}^2 Q_{j,\alpha}^2 \right] \, ,
\eeq
while the system-bath couplings become
\begin{align}
H_{c,L} = -\frac{\kappa_L q_1}{\sqrt{m_L}} \sum_{j=1}^{N_L} (\Lambda_L)_{j,N_L} Q_{j,L} = -\kappa_L q_1 \sum_{j=1}^{N_L} C_{j,L} Q_{j,L} \, , \\
H_{c,R} = -\frac{\kappa_R q_N}{\sqrt{m_R}} \sum_{j=1}^{N_R} (\Lambda_R)_{j,1} Q_{j,L} = -\kappa_R q_N \sum_{j=1}^{N_R} C_{j,R} Q_{j,R} \, ,
\end{align}
where $C_{j,L} = \frac{1}{\sqrt{m_L}} (\Lambda_L)_{j,N_L}$ and $C_{j,R} = \frac{1}{\sqrt{m_R}} (\Lambda_R)_{j,1}$. Finally, we can express the external driving in terms of normal modes:
\beq
H_d(t) = -F(t) \sum_{j=1}^{N_L} D_{j.L} Q_{j,L} \, , 
\eeq
with $D_{j,L} = \frac{1}{\sqrt{m_L}} (\Lambda_L)_{j,1}$.

We now derive the generalized Langevin equations governing the dynamics of the system. The equations of motion for the system are given by
\begin{align}
    \dot{p}_1 &= - \kappa_L q_1 + \kappa_L \sum_{j=1}^{N_L}C_{j,L} Q_{j,L} - U'(q_1) + V'(q_2 -q_1) \, , \label{eq:sys1} \\ 
    \dot{p}_j &=  -U'(q_j) + V'(q_{j+1} - q_j) - V'(q_j - q_{j-1}) \, , \\
    \dot{p}_N& = - \kappa_R q_N + \kappa_R \sum_{j=1}^{N_R}C_{j,R} Q_{j,R} - U'(q_N) - V'(q_N -q_{N-1}) \, , \label{eq:sysN}
\end{align}
where $f'(r)$ denotes the derivative of $f(r)$ with respect to the argument $r$ and $p_j = m \dot{q}_j$. For the reservoirs, we have
\begin{align}
    \ddot{Q}_{j,L} = & - \omega_{j,L}^2 Q_{j,L} + \kappa_L q_1 C_{j,L} + F(t) D_{j,L} \, , \\
    \ddot{Q}_{j,R} = & - \omega_{j,R}^2 Q_{j,R} + \kappa_R q_N C_{j,R} \, .
\end{align}
The bath equations of motion are linear and can be straightforwardly solved given the initial conditions for the reservoir modes $\{Q_{j,\alpha}(t_0), P_{j,\alpha}(t_0)\}$~\cite{weiss2012quantum}:
\begin{align}
    Q_{j,L}(t) & =   Q_{j,L}^{(0)} (t) + \kappa_L \frac{C_{j,L}}{\omega_{j,L}} \int_{t_0}^t ds\,  \sin\left(\omega_{j,L}( t-s )\right) q_1(s) \nonumber \\
    & + \frac{D_{j,L}}{\omega_{j,L}} \int_{t_0}^t ds\,  \sin\left(\omega_{j,L}( t-s )\right) F(s) \, , \label{eq:Lformal} \\
    Q_{j,R}(t) & =   Q_{j,R}^{(0)} (t) + \kappa_R \frac{C_{j,R}}{\omega_{j,R}} \int_{t_0}^t ds\,  \sin\left(\omega_{j,R}( t-s )\right) q_N(s) \, , \label{eq:Rformal}
\end{align}
where $Q_{j,\alpha}^{(0)}$ is the general solution of the homogeneous equation $\ddot{Q}_{j,\alpha} = - \omega_{j,\alpha}^2$ 
\beq
Q_{j,\alpha}^{(0)} = Q_{j,\alpha}(t_0) \cos\left(\omega_{j,\alpha} (t-t_0) \right) + \frac{P_{j,\alpha}(t_0)}{\omega_{j,\alpha}} \sin\left(\omega_{j,\alpha} (t-t_0) \right) \, .
\eeq
Integrating the second term in Eqs.~\eqref{eq:Lformal} and~\eqref{eq:Rformal} by parts and accounting for the initial conditions of the system $q_j(t_0) = 0$, we get
\begin{align}
     Q_{j,L}(t) & =  Q_{j,L}^{(0)} (t) - \kappa_L \frac{C_{j,L}}{\omega_{j,L}^2} \int_{t_0}^t ds\, \cos\left(\omega_{j,L}( t-s )\right) \dot{q}_1(s) \nonumber \\
    & +  \kappa_L \frac{C_{j,L}}{\omega_{j,L}^2} q_1(t) + \frac{D_{j,L}}{\omega_{j,L}} \int_{t_0}^t ds\,  \sin\left(\omega_{j,L}( t-s )\right) F(s) \, , \label{eq:Lformal2} \\
    Q_{j,R}(t) & =   Q_{j,R}^{(0)} (t) - \kappa_R \frac{C_{j,R}}{\omega_{j,R}^2} \int_{t_0}^t ds\,  \cos\left(\omega_{j,R}( t-s )\right) \dot{q}_N(s) \nonumber \\
    & + \kappa_R \frac{C_{j,R}}{\omega_{j,R}^2} q_N(t) \label{eq:Rformal2} \, .
\end{align}
We plug these into the equations of motion of the system: while the ``bulk" degrees of freedom ($j=2,\cdots,N-1$) are unaffected, the ``boundary" equations Eqs.~\eqref{eq:sys1} and Eqs.~\eqref{eq:sysN} become
\begin{align}
\dot{p}_1(t) & =  - U'(q_1) + V'(q_2 - q_1) - \left(\kappa_L - \kappa_L^2 \gamma_L(0) \right) q_1(t) \nonumber \\
& + \xi_L(t) - \int_{t_0}^t ds \gamma_L(t-s) \dot{q}_1(s) + \int_{t_0}^t ds \lambda(t-s) F(s) \, , \\
\dot{p}_N(t) & = -U'(q_N) - V'(q_N - q_{N-1}) -  \left(\kappa_R - \kappa_R^2 \gamma_R(0) \right) q_N(t) \nonumber \\
& + \xi_R(t) - \int_{t_0}^t ds \gamma_R(t-s) \dot{q}_N(s) \, ,
\end{align}
where the quantities
\begin{align}
    \xi_\alpha(t) = & \sum_{j=1}^{N_\alpha} C_{j,L} Q_{j,\alpha}^{(0)}(t) \, \\
    \gamma_\alpha(t) = &\kappa_\alpha^2 \sum_{j=1}^{N_\alpha} \frac{C_{j,\alpha}^2}{\omega_{j,\alpha}^2} \cos\left(\omega_{j,\alpha} t\right) \, , \\
    \lambda(t) = & \kappa_L \sum_{j=1}^{N_L}\frac{C_{j,L} D_{j,L}}{\omega_{j,L}} \sin\left(\omega_{j,L} \right) \, ,
\end{align}
are entirely determined by the spectral properties of the bath.
Observing that 
\begin{align}
\gamma_L(0) = & \sum_{j=1}^{N_L} \frac{C_{j,L}^2}{\omega_{j,L}^2} = (K_L^{-1})_{N_L,N_L} = \kappa_L^{-1} \, ,\nonumber  \\
\gamma_R(0) = & \sum_{j=1}^{N_R} \frac{C_{j,R}^2}{\omega_{j,R}^2} = (K_R^{-1})_{1,1} = \kappa_R^{-1} \, ,
\end{align}
we can further simplify the equations of motion for the system's boundary oscillators,
\begin{align}
\dot{p}_1(t) & =  - U'(q_1) + V'(q_2 - q_1) + \xi_L(t) \nonumber \\
& - \int_{-\infty}^t ds \gamma_L(t-s) \dot{q}_1(s) + \int_{-\infty}^t ds \lambda(t-s) F(s) \, , \label{eq:sysGL1} \\
\dot{p}_N(t) & = -U'(q_N) - V'(q_N - q_{N-1}) + \xi_R(t) \nonumber \\
&  - \int_{-\infty}^t ds \gamma_R(t-s) \dot{q}_N(s) \, , \label{eq:sysGLN}
\end{align}
where we have also taken the limit $t_0 \to -\infty$ since we will be interested in the steady-state properties of the system. 

We recognize Eqs.~\eqref{eq:sysGL1} and~\eqref{eq:sysGLN} as generalized Langevin equations governing the dynamics of the boundary oscillators of the system: $\xi(t)$ is the random forcing term, which is determined by the initial conditions of the reservoir modes; $\gamma(t)$ is the damping kernel, which accounts for the interaction between the system and reservoirs; and $\lambda(t)$ is the force delay kernel~\cite{reichert2016,grabert2018}, which incorporates the effective driving mediated by the bath, which is explicitly being driven. Both $\gamma$ and $\lambda$ are time-retarded and, along with $\xi$, completely characterize the spectral properties of the reservoir. 

With the general form of Langevin equations established, we now take the limit of infinite baths $N_\alpha \to \infty$ as well as the continuum string limit, which corresponds to introducing a lattice spacing $a$ and taking the limits $a\to 0, m_\alpha \to 0, \epsilon_\alpha \to \infty$ while keeping the mass density $\mu_\alpha = m/a$ and elastic modulus $Y_\alpha = \epsilon_\alpha a$ finite. First, as stated earlier, we assume that all oscillators in the reservoir are stationary, so that $Q_{j,\alpha}(t_0) = P_{j,\alpha}(t_0) = 0$ and consequently $\xi_{\alpha}(t) = 0$, resulting in a vanishing random force exerted by the baths. Typically, one assumes that the reservoirs are in thermal equilibrium at some temperature $T_\alpha$, with the initial conditions drawn from the Boltzmann distribution (see e.g., Ref.~\cite{das2012}). Here, however, we ignore finite temperature effects since we are interested in the steady-states that result from the external driving force. 

In order to characterize the damping (or memory) kernel $\gamma_\alpha(t)$, it is conventional to represent it as
\beq
\gamma_\alpha(t) = 2 \kappa_\alpha^2 \int_0^\infty \frac{d\omega}{\pi} \frac{J_\alpha(\omega)}{\omega} \cos(\omega t) \,  ,
\eeq
where $J_\alpha(\omega)$ is the spectral density of bath modes
\beq
J_\alpha(\omega) = \frac{\pi}{2} \sum_{k=1}^{N_\alpha} \frac{C_{k,\alpha}^2}{\omega_{k,\alpha}} \delta(\omega - \omega_{k,\alpha}) \, .
\eeq
In general, the lack of translation invariance in the bath Hamiltonians $H_{B,\alpha}$ makes an analytic calculation of the spectral density somewhat involved. However, it conveniently turns out that the spectral properties of the bath are independent of $\kappa_\alpha$ in the continuum limit in which we are interested~\cite{eckmann1999,das2012,das2020}, so we set $\kappa_\alpha = \epsilon_\alpha$, in which case we can obtain the eigenfunctions and normal mode frequencies of the bath Hamiltonians exactly:
\begin{align}
(\Lambda_\alpha)_{ij} & = \sqrt{\frac{2}{N_\alpha + 1}} \sin \left( \frac{\pi i j}{N_\alpha + 1}\right) \, , \\
\omega_{j,\alpha} & = \omega_{0,\alpha} \sin \left( \frac{\pi j}{2 (N_\alpha + 1)} \right) \, ,
\end{align}
where the Debye frequency $\omega_{0,\alpha} = 2 \sqrt{\epsilon_\alpha/m_\alpha}$ sets the bandwidth of the reservoirs. It is then straightforward to show that 
\begin{align}
\gamma_\alpha(t) = \frac{m_\alpha \omega_{0,\alpha}^2}{2(N_\alpha + 1)} & \sum_{j=1}^{N_\alpha} \left[ \cos^2\left( \frac{\pi j}{2(N_\alpha + 1)} \right) \right. \nonumber \\
& \left. \times \cos\left( \omega_{0,\alpha} t \sin\left( \frac{\pi j}{2 (N_\alpha + 1)} \right) \right) \right]\, .
\end{align}
Taking the limit $N_\alpha \to \infty$, this becomes
\begin{align}
\gamma_\alpha(t) &= \frac{m_\alpha \omega_{0,\alpha}^2}{2} \int_0^\pi \frac{dk}{\pi} \cos^2\left(\frac{k}{2} \right) \cos \left(\omega_{0,\alpha}t \sin\left( \frac{k}{2}\right) \right)  \nonumber \\
& = \frac{m_\alpha \omega_{0,\alpha}}{2} \frac{\mathcal{J}_1 (\omega_{0,\alpha} t)}{t} \, ,
\end{align}
with $\mathcal{J}_{n,x}$ the Bessel function of the first kind. Characteristic of the Rubin model~\cite{rubin1960,rubin1963}, we see that the memory kernel has a memory time $\sim 1/\omega_{0,\alpha}$ set by the bandwidth of the reservoir. In order to obtain a Markovian damping kernel, with a vanishing memory time, we take the limit in which the reservoir is a continuous string
\beq
\gamma_\alpha(t) = \sqrt{m_\alpha \epsilon_\alpha} \frac{\mathcal{J}_1 (\omega_{0,\alpha} t)}{t} \to \gamma_{0,\alpha} \delta(t)
\eeq
where we have taken the limit $m_\alpha \to 0, \epsilon_\alpha \to \infty$ and $\gamma_{0,\alpha} = \sqrt{\mu_\alpha Y_\alpha}$. This corresponds to a bath with an Ohmic spectral density $J_\alpha(\omega) = \gamma_{0,\alpha} \omega$. For the force delay kernel $\lambda(t)$, we similarly find that
\begin{align}
\lambda(t) &= \frac{2 \epsilon_L}{m_L \omega_{0,L}} \frac{4}{N_L + 1} \sum_{j=1}^{N_L}\left[ \sin\left(\frac{\pi j}{2(N_L + 1)} \right) \right. \nonumber \\
& \left. \times \cos\left(\frac{\pi j}{2(N_L + 1)} \right)^2 \sin\left( \omega_{0,L} t \sin\left( \frac{\pi j}{2 (N_\alpha + 1)} \right) \right) \right]
\end{align}
which, in the $N_L \to \infty$ limit takes the integral form
\begin{align}
\lambda(t) &= \frac{2 \epsilon_L}{m_L \omega_{0,L}} \int_0^\pi \frac{dk}{\pi}
\frac{\sin\left( k\right)^2}{\sin\left(\frac{k}{2}\right)^2}
 \sin\left(\omega_{0,L}t \sin\left(\frac{k}{2}\right) \right) \nonumber \\
 & = \frac{8 \epsilon_L}{m_L \omega_{0,L}^2} \frac{\mathcal{J}_2(\omega_{0,L} t)}{t} \, .
\end{align}
In the limit of a continuous string, we again find that the time-retardation vanishes and the force acts instantaneously on the system, such that
\beq
\lambda(t) = 2 \frac{\mathcal{J}_2(\omega_{0,L}t)}{t} \to 2 \delta(t) \, .
\eeq

Thus, starting from a microscopic model of the reservoirs, we have derived the following Langevin equations for the system's degrees of freedom:
\begin{align}
    \dot{p}_1 &= - U'(q_1) + V'(q_2 -q_1) -\gamma_{0,L} \dot{q}_1(t) + 2 F(t) \,   \label{eq:LE1App} \\ 
    \dot{p}_j &=  -U'(q_j) + V'(q_{j+1} - q_j) - V'(q_j - q_{j-1}) \,  \\
    \dot{p}_N& = - U'(q_N) - V'(q_N -q_{N-1}) - \gamma_{0,R} \dot{q}_N(t) \, , \label{eq:LEnApp} 
\end{align}
where we have taken the limit of semi-infinite reservoirs, which are at zero temperature and are taken to be continuous strings. We see that these are precisely the equations of motion of a finite chain of anharmonic oscillators with Hamiltonian dynamics governed by $H_S$ in Eq.~\eqref{eq:sysHam}, which is driven by an external force $F(t)$ at its left boundary and experiences damping at both ends (see Fig.~\ref{fig:setup2}).

\end{document}